\documentclass[%
  twoside,
  reprint,
  amsmath,amssymb,
  aps,
  pra,
  nofootinbib,
  showpacs,
  superscriptaddress,
  a4paper
]{revtex4-1}

\usepackage{amsmath}
\usepackage{amssymb,upref}
\usepackage{anyfontsize}
\usepackage[usenames,dvipsnames]{xcolor}
\usepackage{graphicx}
\usepackage{subfigure}

\usepackage[utf8]{inputenc}
\usepackage[T1]{fontenc}

\usepackage{lipsum}

\usepackage{siunitx}  

\usepackage[normalem]{ulem}

\graphicspath{{./},{./Figures/}}

\usepackage[
centering, includefoot,
text={7.1in,10in},
total={6.3in,8.75in},
top=1in, left=0.62in,
]{geometry}

\makeatletter
\renewcommand\frontmatter@abstractwidth{\dimexpr\textwidth-1.2in\relax}
\makeatother

\usepackage[%
  bookmarks=true,
  colorlinks,
  linkcolor=blue,
  urlcolor=blue,
  citecolor=blue,
  plainpages=false,
  pdfpagelabels,
  final,
  breaklinks=true
]{hyperref}
\hypersetup{
pdftitle={	Strong-field approximation in a rotating frame: High-order harmonic emission from p states in bicircular fields}, 
pdfauthor={Emilio Pisanty and Álvaro Jiménez-Galán},
pdfkeywords={strong-field physics, ultrafast physics, high-harmonic generation, high-order harmonic generation, bicircular fields, imaginary time method, quantum orbits, quantum physics, quantum mechanics, atomic physics, atomic structure, strong-field approximation, SFA}
}

\usepackage{natbib}

\makeatletter \def\NAT@def@citea{\def\@citea{\NAT@separator\,}} \makeatother

\newcommand{\reffig}[1]{Fig.~\ref{#1}}

\newcommand{\citenisteq}[1]{\cite[Eq.\,(\href{http://dlmf.nist.gov/#1}{#1})]{NIST_handbook}}

\DeclareFontFamily{U} {MnSymbolA}{}
\DeclareFontShape{U}{MnSymbolA}{m}{n}{
  <-6> MnSymbolA5
  <6-7> MnSymbolA6
  <7-8> MnSymbolA7
  <8-9> MnSymbolA8
  <9-10> MnSymbolA9
  <10-12> MnSymbolA10
  <12-> MnSymbolA12}{}
\DeclareSymbolFont{MnSyA} {U} {MnSymbolA}{m}{n}
\DeclareMathSymbol{\lcirclearrowright}{\mathrel}{MnSyA}{252}
\DeclareMathSymbol{\rcirclearrowleft}{\mathrel}{MnSyA}{250}

\usepackage{physics}
\renewcommand{\d}{\ensuremath{\textrm{d}}}
\renewcommand{\Re}{\operatorname{Re}}
\renewcommand{\Im}{\operatorname{Im}}
  \newcommand{\vbr}{\vb{r}}
  \newcommand{\vba}{\vb{A}}
  \newcommand{\vbf}{\vb{F}}
  \newcommand{\vbp}{\vb{p}}
  \newcommand{\vbk}{\vb{k}}
  \newcommand{\vbd}{\vb{d}}
  \newcommand{\vbD}{\vb{D}}
  \newcommand{\bsa}{\boldsymbol{\alpha}}
  
\newcommand{\rightpol}{\!\; \!\!\rcirclearrowleft}
\newcommand{\leftpol }{\!\; \!\!\lcirclearrowright}

\newcommand{\ue}[1]{\hat{\vb{e}}_{#1}}

\newcommand{\Lz}{\hat{L}_z}

\newcommand{\rot}{\mathsf{R}}
\newcommand{\lab}{\mathsf{L}}
\newcommand{\las}{\mathsf{las}}
\newcommand{\tref}{{t_\mathrm{ref}}}

\newcommand{\volkov}[2]{\ket{\Psi_{#1}(#2)}}
\newcommand{\valpha}[2]{\ket{\Psi^{(\alpha)}_{#1}(#2)}}

\newcommand{\valphasm}[2]{\ket*{\Psi^{(\alpha)}_{#1}(#2)}}
\newcommand{\valphabsm}[2]{\bra*{\Psi^{(\alpha)}_{#1}(#2)}}

\newcommand{\cc}{\ensuremath{\mathrm{c.c.}}}
\newcommand{\eps}{\varepsilon}
\newcommand{\Y}{\Upsilon}

\begin{document}

\title{Strong-field approximation in a rotating frame: high-order harmonic emission\\ from \textit{p} states in bicircular fields}

\author{Emilio Pisanty}
 \email{emilio.pisanty@icfo.eu}
 \affiliation{Max-Born Institute for Nonlinear Optics and Short Pulse Spectroscopy, Berlin, Germany}
 \affiliation{ICFO -- Institut de Ciències Fotòniques, The Barcelona Institute of Science and Technology, 08860 Castelldefels (Barcelona), Spain}
\author{Álvaro Jiménez-Galán}
 \email{jimenez@mbi-berlin.de}
 \affiliation{Max-Born Institute for Nonlinear Optics and Short Pulse Spectroscopy, Berlin, Germany}
\date{December 1, 2017}

\begin{abstract}
High-order harmonic generation with bicircular fields -- the combination of counter-rotating circularly polarized pulses at different frequencies -- results in a series of short-wavelength XUV harmonics with alternating circular polarizations, and experiments show that there is an asymmetry in the emission between the two helicities: a slight one in helium, and a larger one in neon and argon, where the emission is carried out by \textit{p}-shell electrons.
Here we analyze this asymmetry by switching to a rotating frame in which the field is linearly polarized; this induces an effective magnetic field which lowers the ionization potential of the \textit{p}${{}_{+}}\!$ orbital that co-rotates with the lower-frequency driver, enhancing its harmonic emission and the overall helicity of the generated harmonics, while also introducing nontrivial effects from the transformation to a non-inertial frame in complex time. 
In addition, this analysis directly relates the small asymmetry produced by \textit{s}-shell emission to the imaginary part of the recollision velocity in the standard strong-field-approximation formalism.
\\[-2mm]

\noindent
\footnotesize
Accepted manuscript for
\href{%
  https://doi.org/10.1103/PhysRevLett.122.203201%
  }{%
  \color[rgb]{0,0,0.55}%
  \textit{Phys.\ Rev.\ A} \textbf{96}, 063401 (2017)%
  }. 
Available as %
\href{%
  https://arxiv.org/abs/1709.00397%
  }{%
  \color[rgb]{0,0,0.55}%
  arXiv:1709.00397%
  } %
under %
\href{%
  %
  https://creativecommons.org/licenses/by-nc-sa/4.0/
  }{%
  \color[rgb]{0,0,0.55}%
  CC BY-NC-SA%
  }.
\\[-6mm]

\end{abstract}

\maketitle

Light is one of our main tools for the investigation of the internal structure of dynamics of matter, and in this role we employ all of its characteristics: its spatial and temporal aspects, its coherence as a wave phenomenon, and its polarization as a vector effect. As we probe deeper into the structures of materials and molecules, and as we look with increasing detail at their dynamics, it becomes necessary to use higher frequencies and shorter pulses, and here the process of high-order harmonic generation (HHG) stands out as a simple and effective way to produce bright, short, coherent pulses of high-frequency radiation~\cite{krausz-ivanov_attosecond-review_2009}. 

The process of HHG is essentially driven by the ionization of gases by a strong, long-wave\-length laser pulse, which then drives the photoelectron back to its parent ion with a high energy, which it emits as a single photon. This permits a large flexibility in the emission process, and its sub-laser-cycle nature allows us to probe atomic and molecular systems at their own timescales. Unfortunately, however, its collision-driven nature has long left unavailable one of the crucial tools in the toolbox -- the use of circular polarizations~\cite{budil-ellipticity-1993}.

A number of attempts have been made over the years to produce high-order harmonics with circular or elliptical polarization~\cite{vodungbo_hhg-waveplate_2011, mairesse_high-harmonic-spectroscopy_2010, stremoukhov_elliptical-harmonics_2016, strelkov_harmonics-polarization-origin_2011, ferre_circular-harmonics_2015, pisanty_PhD-thesis_2016}, which would enable detailed time-resolved studies of magnetic materials and chiral molecules, but they have generally suffered from complex configurations, low efficiencies, and limited harmonic ellipticities. These limitations were recently overcome by combining counter-rotating circularly polarized drivers~\cite{ fleischer_spin_2014,kfir_generation_2015}, in a so-called `bicircular' configuration; this produces fully circular harmonics at similar efficiency to linear-polarization schemes, and it can be implemented with minimal modifications to existing beamlines~\cite{kfir_in-line-bicircular_2016}.

The simplest configuration uses drivers with equal intensities at frequencies $\omega_1=\omega$ and \mbox{$\omega_2=2\omega$}, in which case the fields combine to make a trefoil-shaped Lissajous figure~\cite{ivanov_nature_photonics_2014}, with a three-fold dynamic symmetry: the system is unchanged under the combined action of a rotation by $\SI{120}{\degree}$ and a temporal delay by one third of the period of the fundamental. This dynamical symmetry enforces a corresponding selection rule on the harmonic response of the system~\cite{alon-averbukh_selection-rules_1998}, which only permits the emission of harmonics at frequency $(3n+1)\omega$, with the same polarization as the fundamental (right-handed, $\rightpol$), and at frequency $(3n-1)\omega$, with the same polarization as the second harmonic (left-handed, $\leftpol$).

This selection rule, however, is silent on the relative strength of the harmonic emission at these two helicities: it specifies what \textit{can} happen, but not the amplitude at which it does. It therefore came as a surprise when experiments showed a definite asymmetry between the two helicities in the plateau harmonics, with a preference for right-handed harmonics that co-rotate with the lower-frequency driver~\cite{ fan_bright-circularly_2015, kfir_generation_2015, baykusheva_bicircular-hhg-spectroscopy}. There is some debate over the origin of this asymmetry, since it can have a macroscopic origin from chiral phase-matching properties~\cite{kfir_chiral-phase-matching_2016}; on the other hand, it is also present in numerical simulations of the time-dependent Schrödinger equation~\cite{ baykusheva_bicircular-hhg-spectroscopy, medisauskas_generating_2016, zhang_helicity-reversion_2016, jimenez_control-of-polarization_2017}, so it appears to come from both microscopic and macroscopic effects. 

More intriguingly, the asymmetry is much stronger in neon than in helium, so it is evidently caused by (and a good testing ground for) the harmonic emission from $p$-shell electrons~\cite{milosevic_circularly_2015}; this makes it an object of intrinsic interest, since the contributions of the orbital angular momentum of electrons to HHG emission are relatively hard to bring to the fore. Moreover, this asymmetry is technologically relevant, since an asymmetric spectrum is more chiral~\cite{dorney_helicity_2017}, and is therefore less dependent on spectral filtering for its use in chiral experiments like enantiomer detection~\cite{lux-baumert_PECD_2012} or x-ray magnetic circular dichroism~\cite{kfir_generation_2015}, so it can be applied even in systems with a broad spectral response.

\begin{figure}[htbp]
\centering
\begin{tabular}{c}
\subfigure{\label{fig-fields-figure-lab-frame}%
\includegraphics[scale=1]{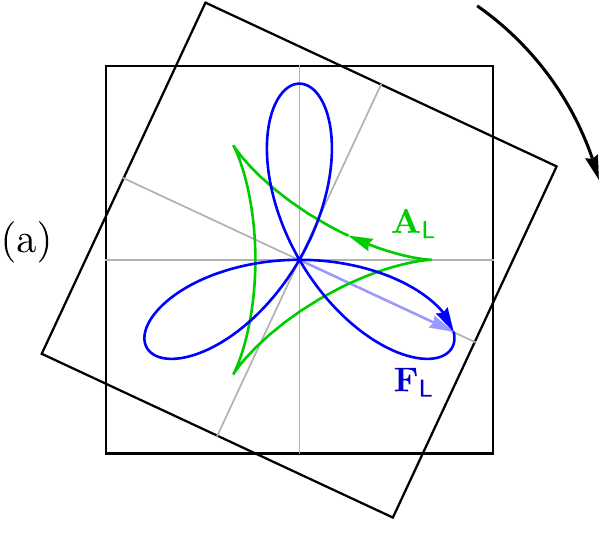}}\\
\subfigure{\label{fig-fields-figure-rotating-frame}%
\includegraphics[scale=1]{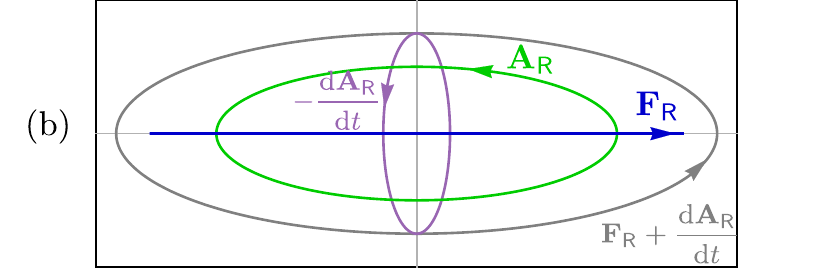}}
\end{tabular}
\caption{%
(\hyperref[fig-fields-figure-lab-frame]{a}) Electric field and vector potential in the laboratory frame, and transformation to the rotating frame.
(\hyperref[fig-fields-figure-rotating-frame]{b})
Fields on the rotating frame: electric field $\vbf_\rot$ (blue), `vector potential' $\vba_\rot$ (green), the derivative of the vector potential $-\frac{\d\vba_\rot}{\d t}$ (purple) and the difference $\vbf_\rot+\frac{\d\vba_\rot}{\d t}=\bsa \times\vba_\rot(t)$ of the latter with the electric field (gray).}
\label{fig:fields-figure}
\end{figure}

In this paper we examine the asymmetry in the emission of opposite helicities by analyzing the harmonic generation process in a non-inertial frame which rotates at half the frequency difference between the two drivers, $\alpha=\tfrac12(\omega_2-\omega_1)=\tfrac12 \omega$~\cite{zuo_hhg-magnetic_1995, mauger_circularly_2016}, as explored in depth for bicircular HHG by Reich and Madsen~\cite{reich-madsen_rotating-frame-bicircular_2016,reich-madsen_molecular-symmetries_2016}, in which frame the bicircular fields become monochromatic and linearly polarized, as shown in \reffig{fig:fields-figure}. %
We work within the workhorse Strong-Field Approximation (SFA)~\cite{LewensteinHHG, HHGTutorial}, which is well understood for bicircular fields in the laboratory frame~\cite{pisanty_PhD-thesis_2016, milosevic-becker-kopold_coplanar-field-mixing_2000}, and which explains the harmonic emission in terms of complex-valued quantum trajectories~\cite{ayuso_attosecond_2017, ivanov-anatomy-2006}.

The transformation to the rotating frame induces two main effects as regards the dynamics. On one hand, the frame's rotation introduces a Coriolis term of the form $\alpha \Lz$ in the hamiltonian; this is an effective magnetic field and it shifts the ionization potential of the $p$-shell states, thereby altering their contribution to the ionization. As a much less intuitive effect, the frame transformation $R_z(\alpha t)$ involves a rotation by complex angles when the ionization time is imaginary, and this changes the trajectories in ways that are interpreted differently by the two $p$-state transition matrix elements. In this work we examine in detail these effects, which bring to the fore the role of complex times in the ionization dynamics driven by circularly polarized fields~\cite{barth-smirnova_nonadiabatic-circular_2017}, and more generally, give an interesting window into the behaviour of complex-time methods in non-inertial frames.

In addition, since in the rotating frame the driver is monochromatic, its harmonic emission forms a single comb, with both elements of each line pair superposed, and this enables us to study the line-pair emission asymmetry through the polarization of each rotating-frame harmonic. This provides a new perspective into the quantum-orbit dynamics, and it allows us to conclusively tie the asymmetry in the $s$-state emission to a nonzero imaginary component of the recollision velocity in bicircular fields, in both the rotating and the laboratory frames.

This paper is structured as follows. In section~\ref{sec-sfa-in-a-rotating-frame} we formulate the strong-field approximation in the rotating frame, and we present the resulting spectra and polarization of the harmonics in section~\ref{sec-harmonic-spectra}. Then, in section~\ref{discussion}, we analyze the origin of the helicity asymmetry, and its grounding in the quantum-orbit trajectories in the rotating frame. We include, in appendix~\ref{sec-appendix-lab-frame}, laboratory-frame versions of our rotating-frame results, for comparison, and in appendix~\ref{sec-appendix-matrix-elements} we derive the necessary $p$-state dipole transition matrix elements.

\section{The strong-field approximation in a rotating frame}
\label{sec-sfa-in-a-rotating-frame}
We consider the hamiltonian for the interaction of a single electron with a strong laser field in the length gauge, of the form
\begin{equation}
H_\lab(t)=\frac12 \hat{\vbp}^2 + V_0(\hat\vbr) +\hat{\vbr}\cdot \vbf(t)
,
\label{initial-lab-frame-hamiltonian}
\end{equation}
where $\vbf(t)$ is the external field, $V_0(\hat\vbr)$ is an effective atomic potential, and we use atomic units unless otherwise stated. Here we have taken the single-active-electron (SAE) approximation, as in previous works~\cite{medisauskas_generating_2016}, in the understanding that one should calculate the emission dipoles from all the participating orbitals and then add them coherently (as dipole vectors, but incoherently with respect to the single-electron Hilbert space~\cite{milosevic_circularly_2015}). The external field we take in the form
\begin{equation}
\vbf(t)
=
\Re\left(
F \ue 1 e^{-i\omega t}
+
F \ue 2 e^{-2i\omega t}
\right)
,
\label{initial-external-field}
\end{equation}
where $\ue1=\ue+$ and $\ue 2=\ue -$ are the circular-polarization unit vectors $\ue\pm=\mp(\ue x\pm i\ue y)/\sqrt{2}$.

To analyze this hamiltonian we change to the rotating frame~\cite{reich-madsen_rotating-frame-bicircular_2016, reich-madsen_molecular-symmetries_2016, zuo_hhg-magnetic_1995, mauger_circularly_2016} through the unitary transformation $U(t)=e^{-i\alpha t \Lz}$, which affects states as $\ket{\psi_\lab} \mapsto \ket{\psi_\rot}=U(t)\ket{\psi_\lab}$ and vector operators such as the position operator via
\begin{equation}
U(t) \hat{\vbr} U(t)^\dagger
=
R_z(\alpha t) \hat{\vbr}
=
\begin{pmatrix}
 \cos(\alpha t) & \sin(\alpha t) & 0 \\
-\sin(\alpha t) & \cos(\alpha t) & 0 \\
0 & 0 & 1
\end{pmatrix}
\begin{pmatrix}
\hat x \\ \hat y \\ \hat z
\end{pmatrix}
.
\end{equation}
Similarly, the lab-frame hamiltonian $H_\lab(t)$ is translated to the rotating-frame hamiltonian $H_\rot(t)$ through
\begin{align}
\hat H_{\rot}(t)
& =
U(t) \hat H_\lab(t) U(t)^\dagger - iU(t) \frac{\partial}{\partial t}U(t)^\dagger
\nonumber \\ & =
\frac12 \hat{\vbp}^2 + V_0(\hat\vbr) +\hat{\vbr}\cdot \vbf_\rot(t) + \alpha \Lz
,
\end{align}
where we have assumed a spherically symmetric atomic potential $V_0(\vbr)$, and we use the fact that $M\vb{u}\cdot \vb{v} = \vb{u} \cdot M^T \vb{v}= \vb{u}\cdot M^{-1}\vb{v}$ for any orthogonal matrix $M$, to obtain the rotating-frame electric field $\vbf_\rot(t) = R_z^{-1}(\alpha t)\vbf(t)$.


Generally speaking, the effect of the frame transformation on time-dependent vectors is to blueshift right-circular fields by $\alpha$, and to similarly redshift left-circular fields. This comes from the action of the rotation matrix on the circular unit vectors $\ue\pm$
\begin{align}
R_z^{-1}(\alpha t) \ue\pm
& =
\mp \frac{1}{\sqrt{2}} R_z^{-1}(\alpha t)(\ue x\pm i\ue y)
 =e^{\mp i\alpha t} \ue\pm
 ,
\label{rotation-of-unit-circular-vectors}
\end{align}
which therefore means that a circularly-polarized vector $\vb{a}(t) = \Re(a\,\ue\pm e^{-i\nu t})$ transforms to 
\begin{align}
\vb{a}_\rot(t) 
& = 
R_z^{-1}(\alpha t) \vb{a}(t)
 = 
\Re(ae^{-i\nu t}\,R_z^{-1}(\alpha t) \ue \pm )
\nonumber \\ & = 
\Re(ae^{-i(\nu\pm\alpha) t} \ue \pm )
.
\end{align}
For the driving field in~\eqref{initial-external-field}, taking $\alpha=\omega/2$ shifts the two components towards a single frequency, giving a linearly polarized field
\begin{equation}
\vbf_\rot(t) = F\cos(\frac32 \omega t)\ue x.
\label{rotating-frame-f}
\end{equation}
The field in~\eqref{rotating-frame-f} can only emit odd harmonics of $\frac32\omega$ in the rotating frame, but the circular components of the $(2k+1)$th harmonic will then be shifted to $(2k+1)\frac32\omega \pm \frac12\omega=\left(3k+\frac{3\pm1}{2}\right)\omega$, so we recover the lab-frame selection rules.


Having transformed the full Hamiltonian to the rotating frame, we now proceed in the fashion of the standard Strong-Field Approximation (SFA) formalism \cite{LewensteinHHG, HHGTutorial}, which propagates the wavefunction under the action of the atomic hamiltonian,
\begin{align}
\hat H_{0,\rot}
& =
\frac12 \hat{\vbp}^2 + V_0(\hat\vbr) + \alpha \Lz
,
\end{align}
until the time of interaction with the laser (or ionization time), $t'$; the dynamics is governed from then onwards by the laser hamiltonian
\begin{align}
\hat H_{\rot,\las}(t)
& =
\frac12 \hat{\vbp}^2 + \alpha \Lz + \hat{\vbr}\cdot \vbf_\rot(t)
\nonumber \\ & =
\frac12 \hat{\vbp}^2 + \alpha \Lz + \hat{V}_{\rot,\las}(t)
.
\label{rotating-frame-laser-hamiltonian}
\end{align}

The wavefunction in the rotating frame thus reads
\begin{align}
\ket{\Psi(t)}
& =
-i \int_{\tref}^t\d t'
U_{\rot,\las}(t,t') \hat{V}_{\rot,\las}(t') U_{0,\rot}(t',\tref)\ket{g}
\nonumber \\ & \qquad \qquad + 
U_{0,\rot}(t,\tref)\ket{g}
,
\end{align}
where the $U_\alpha(t_1,t_2)=\mathrm{T}\{\exp[-i \int_{t_1}^{t_2} H_\alpha\,dt]\}$ are the atomic and laser-driven time-ordered evolution operators. Our main observable will be the time-dependent dipole moment in the rotating frame, which is then of the form
\begin{align}
\vbD_\rot(t)
& = 
\matrixel{\Psi(t)}{-\hat{\vbr}}{\Psi(t)}
\nonumber \\ & \approx
i
\bra{g}
U_{0,\rot}(\tref,t)
\hat{\vbr}
\int_{\tref}^t\d t'
U_{\rot,\las}(t,t')
\nonumber \\ & \qquad \qquad \quad \times
\hat{V}_{\rot,\las}(t') 
U_{0,\rot}(t',\tref)
\ket{g}
+\cc
\label{time-dependent-dipole-explicit}
\end{align}
once we neglect continuum-continuum transitions; here $+\cc$ represents the complex conjugate of the previous term, which we will drop unless necessary. The time-dependent dipole in the rotating frame presents two key modifications with respect to that calculated in the laboratory frame:
\begin{itemize}

\item
The addition of the Coriolis term $\alpha \hat{L}_z$, modifies the ionization potential differently for different $p$ orbitals:
\begin{equation}
\hat H_{0,\rot} \ket{p_\pm}
=
\left[-I_p \pm \alpha \right] \ket{p_\pm}
,
\end{equation}
leading to a higher ionization rate for the $p_+$ orbital~\cite{perelomov-ionization-1966, ammosov-delone-krainov-1986}.
\item
The Schrödinger-equation solutions for the hamiltonian in (\ref{rotating-frame-laser-hamiltonian}) are now the rotating-frame Volkov states. To calculate them, we start by their usual definition in the laboratory frame~\cite{bergou_volkov-wavefunctions-I_1980},
\begin{equation}
\volkov \vbp t
=
e^{-\frac i2 \int^t_{t'} (\vbp + \vba(\tau))^2 \d \tau}
\ket{\vbp+\vba(t)}
,
\end{equation}
where $\vba(t)$ is the vector potential of the field, satisfying $\vbf(t) = -\frac{\d\vba}{\d t}$ and $\ket{\vbp+\vba(t)}$ is a plane wave with kinetic momentum $\vbp+\vba(t)$; this Volkov state obeys the Schrödinger equation
\begin{equation}
i\partial_t \volkov \vbp t
=
\left[
\frac12 \hat{\vbp}^2 + \hat{\vbr}\cdot \vbf(t)
\right]
\volkov \vbp t
.
\end{equation}
These states are easiest to understand by considering the temporal evolution of the plane-wave component on its own: this obeys a Schrödinger equation of the form
\begin{equation}
i\partial_t \ket{\vbp + \vba(t)}
=
\hat{\vbr}\cdot\vbf(t) \ket{\vbp + \vba(t)},
\end{equation}
and since the solutions remain as plane-wave eigenstates of the kinetic energy, the addition of the kinetic phase $e^{-\frac i2 \int^t_{t'} (\vbp + \vba(\tau))^2 \d \tau}$ is a trivial step.

The transformation of the plane-wave component into the rotating frame is then simple to implement, since we only need to rotate the eigenvalue,
\begin{equation}
U(t) \ket{\vbp + \vba(t)}
=
\ket{R_z^{-1}(\alpha t)(\vbp+\vba(t))}
,
\end{equation}
and it is easy to show directly that this state obeys the correct Schrödinger equation,
\begin{align}
i\partial_t \ket{R_z^{-1}(\alpha t)(\vbp+\vba(t))}
& =
\left[
\hat{\vbr}\cdot \vbf_\rot(t)
+
\alpha\Lz
\right]
\nonumber \\ & \quad \cdot
\ket{R_z^{-1}(\alpha t)(\vbp+\vba(t))}
.
\end{align}
Since the solution remains as a plane wave for all time, we can simply add the kinetic phase directly, to obtain the rotating-frame Volkov states,
\begin{equation}
\valpha{\vbp}{t}
=
e^{-\frac i2 \int^t_{t'} (\vbp + \vba(\tau))^2 \d \tau}
\ket{R_z^{-1}(\alpha t)(\vbp+\vba(t))}
,
\end{equation}
which are the rotating-frame continuum solutions of the Schrödinger equation under the hamiltonian in~\eqref{rotating-frame-laser-hamiltonian}.

\end{itemize}

We can now add in the known dynamics of our ground state and the continuum, through the relations $ U_{0,\rot}(t,\tref)\ket{g} = e^{i(I_p-m\alpha)(t-\tref)}\ket{g}$, where $m$ is the magnetic quantum number of the ground state, i.e. $\Lz\ket g=m\ket g$, and the  laser propagator in the form $U_{\rot,\las}(t,t')=\int \d\vbp \valphasm{\vbp}{t}\valphabsm{\vbp}{t'}$, which turns the harmonic dipole into
\begin{align}
\vbD_\rot (t)
& = 
i
\int_{\tref}^t \!\!\! \d t' \!
\int \!\! \d\vbp
\,\,
\vbd^*
  \mathopen{}\left(
  R_z^{-1}(\alpha t)(\vbp+\vba(t))
  \right)\mathclose{}
\,
\nonumber \\ & \qquad \quad \times
\Y\mathopen{}\left(
  R_z^{-1}(\alpha t')(\vbp+\vba(t'))
\right)\mathclose{}
\nonumber \\ & \qquad \quad \times
e^{
   -i(I_p-m\alpha)(t-t')
   -\frac i2 \int^t_{t'} (\vbp + \vba(\tau))^2 \d \tau
   }
.
\end{align}
In the above, $\vbd(\vbk) = \matrixel{\vbk}{\hat{\vbr}}{g}$ is the dipole transition matrix element between the ground state and a plane wave, and we have reduced the ionization dipole via integration by parts to the momentum-space ground state wavefunction $\Y(\vbk) = \left(\frac12\vbk^2+I_p\right)\braket{\vbk}{g}$ following~\cite{HHGTutorial}.

To fully specialize the analysis to the rotating frame, we now perform a change of variables of the form $\vbp \mapsto \vbp_\rot=R_z^{-1}(\alpha t')\vbp$, for each of the momentum integrals as indexed by the ionization time $t'$, giving us a harmonic dipole in the form
\begin{align}
\vbD_\rot (t)
& = 
i
\int_{\tref}^t \!\!\! \d t' \!
\int \!\!\! \d\vbp_\rot
\,\,
\vbd^*
  \mathopen{}\left(
  R_z^{-1}(\alpha (t-t'))\vbp_\rot+\vba_\rot(t)
  \right)\mathclose{}
\,
\nonumber \\ & \quad \times
\Y\mathopen{}\left(
  \vbp_\rot+\vba_\rot(t')\right)
\mathclose{}
\nonumber \\ & \quad \times
e^{
   -i(I_p-m\alpha)(t-t')
   -\frac i2 \int^t_{t'} (R_z^{-1}(\alpha (\tau-t'))\vbp_\rot + \vba_\rot(\tau))^2 \d \tau
   }
.
\end{align}
This then changes the ionization matrix element to a single kinetic momentum, $\vbp_\rot+\vba_\rot(t')$, and it also ensures that the canonical momentum has been rotated via $R_z^{-1}(\alpha (t-t'))$ by the time of recollision through the action of the Coriolis force.

Most importantly, the role of the vector potential $\vba(t)$ in the SFA expression is now taken by the rotating-frame potential
\begin{equation}
\vba_\rot(t) = R_z^{-1}(\alpha t)\vba(t),
\end{equation}
which is a significant change since this is no longer a true vector potential, because its time derivative no longer coincides with the electric field vector in the rotating frame. Instead, we have
\begin{align}
\vbf_\rot(t)
& =
-\frac{\d\vba_\rot}{\d t}(t)
+
\bsa \times\vba_\rot(t)
,
\end{align}
which is the standard connection between time derivatives in the laboratory and rotating frames~\cite{lanczos-variational-principles}, where $\bsa=\alpha \ue z$ is the rotation axis and the cross product $\bsa{\times}$, seen as a linear operator, is cleanly related to the derivative of the rotation matrix as
\begin{align}
\bsa{\times}
& =
-\alpha R_z'(\theta)R_z^{-1}(\theta)
=
-\alpha R_z^{-1}(\theta) R_z'(\theta)
\nonumber \\ & 
=
\alpha 
\begin{pmatrix}
0 & -1 & 0 \\
1 & 0 & 0 \\
0 & 0 & 0
\end{pmatrix}
.
\end{align}

This means, then, that the rotating-frame vector potential is no longer linearly polarized, as a simple consequence of the frequency shifts of circular fields in~\eqref{rotation-of-unit-circular-vectors}: at equal intensities, the contribution of the second harmonic to $\vba(t)$ is reduced by a factor of two, so once the two components are shifted to the same frequency, the total field is elliptically polarized, as shown in \reffig{fig:fields-figure}. This effect increases with the frequency difference, so in $\omega:3\omega$ and higher-order schemes, the rotating-frame vector potential, which determines the SFA action and therefore the corresponding quantum-orbit trajectories, is even closer to circular.

\begin{widetext}

To conclude our manipulations of the harmonic dipole, it is worth performing an explicit saddle-point analysis over the momentum integration, which has the action
\begin{align}
S(\vbp_\rot,t,t') 
& = 
(I_p-m\alpha)(t-t') 
%
+ 
\frac 12 \int^t_{t'} (R_z^{-1}(\alpha (\tau-t'))\vbp_\rot + \vba_\rot(\tau))^2 \d \tau
.
\end{align}
Since the action is quadratic in $\vbp_\rot$, there is a unique solution of the return equation $\frac{\partial S}{\partial\vbp_\rot}=0$, which has the form
\begin{align}
\vbp_{\rot,s}(t,t')
& =
\frac{-1}{t-t'}
\int^t_{t'}R_z(\alpha (\tau-t'))\vba_\rot(\tau)  \, \d \tau
,
\end{align}
with the rotation again caused by the Coriolis effect on the rotating frame.

Similarly, performing the saddle-point approximation \cite{BruijnAsymptotics} with respect to the momentum integration therefore gives
\begin{align}
\label{final-harmonic-dipole}
\vbD_\rot (t)
& = 
i
\int_{\tref}^t \!\!\! \d t' \,
\left(\frac{2\pi}{\eps+i(t-t')}\right)^{3/2}
\vbd^*
  \mathopen{}\left(
  R_z^{-1}(\alpha (t-t'))\vbp_{\rot,s}(t,t')+\vba_\rot(t)
  \right)\mathclose{}
\,
\Y\mathopen{}\left(
  \vbp_{\rot,s}(t,t')+\vba_\rot(t')
\right)\mathclose{}
%
\nonumber \\ & \qquad \qquad \qquad \qquad \quad \times
e^{
   -i(I_p-m\alpha)(t-t')
   -\frac i2 \int^t_{t'} (R_z^{-1}(\alpha (\tau-t'))\vbp_{\rot,s}(t,t') + \vba_\rot(\tau))^2 \d \tau
   }
,
\end{align}
where the added factor represents the wavepacket diffusion over time $t-t'$, with an added regularization factor $\eps$ coming from a failure of the momentum saddle-point approximation at $t-t'\ll 1/I_p$. Finally, to connect the time-dependent harmonic dipole to the experimental spectra we take the Fourier transform,
\begin{align}
\vbD_\rot (\Omega)
& = 
i
\int_{-\infty}^\infty\!\!\! \d t \!\!
\int_{\tref}^t \!\!\! \d t' \,
\left(\frac{2\pi}{\eps+i(t-t')}\right)^{3/2}
\vbd^*
  \mathopen{}\left(
  R_z^{-1}(\alpha (t-t'))\vbp_{\rot,s}(t,t')+\vba_\rot(t)
  \right)\mathclose{}
\,
\Y\mathopen{}\left(
  \vbp_{\rot,s}(t,t')+\vba_\rot(t')
\right)\mathclose{}
\nonumber \\ & \qquad \qquad \qquad \qquad \quad \times
e^{
   +i\Omega t
   -i(I_p-m\alpha)(t-t')
   -\frac i2 \int^t_{t'} (R_z^{-1}(\alpha (\tau-t'))\vbp_{\rot,s}(t,t') + \vba_\rot(\tau))^2 \d \tau
   }
,
\label{final-harmonic-dipole-frequency-domain}
\end{align}
and we analyze this double temporal integral using the standard saddle-point methods~\cite{ LewensteinHHG, HHGTutorial, milosevic-becker-kopold_coplanar-field-mixing_2000, BruijnAsymptotics}, giving a sum of contributions coming from discrete times $t_s,t_s'$ which represent discrete quantum orbits. Our implementation is available from Refs.~\citealp{RB-SFA, FigureMaker}.
\end{widetext}

$\ $

$\ $

\begin{figure*}[t!]
\centering
\begin{tabular}{c}
\subfigure{ \label{fig-single-burst-intensity-ho}
\includegraphics[scale=1]{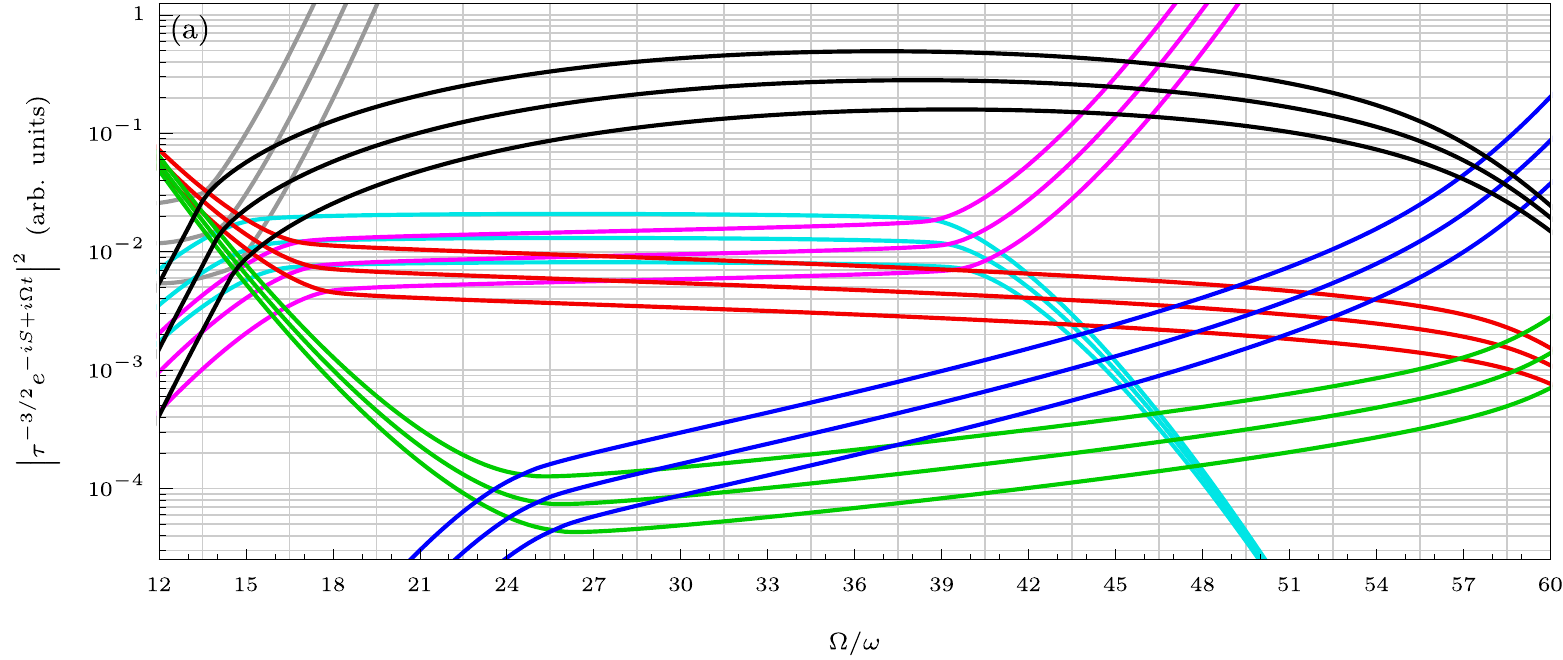}}\\
\subfigure{ \label{fig-single-burst-tau-ho}
\includegraphics[scale=1]{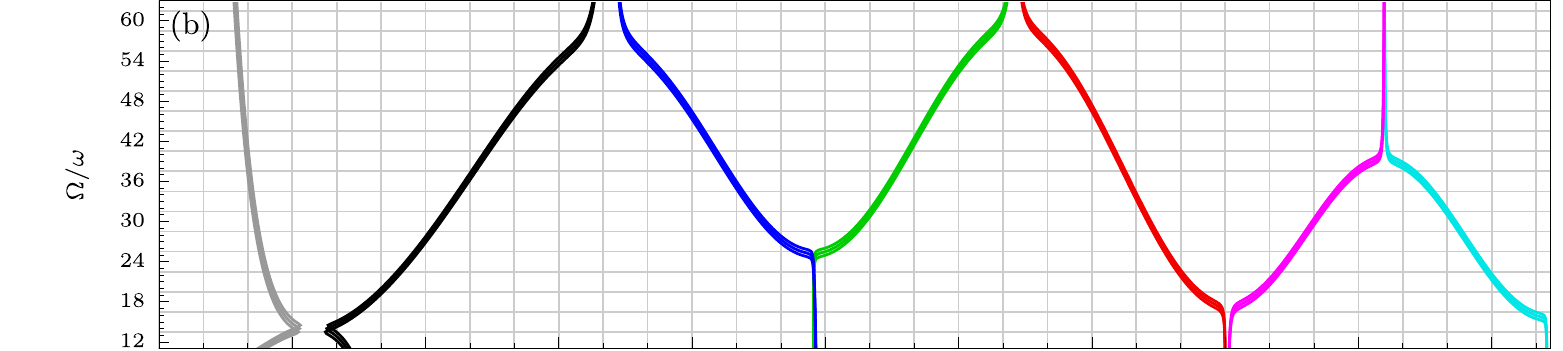}}\\[-2mm]
\subfigure{ \label{fig-single-burst-int-ttau}
\includegraphics[scale=1]{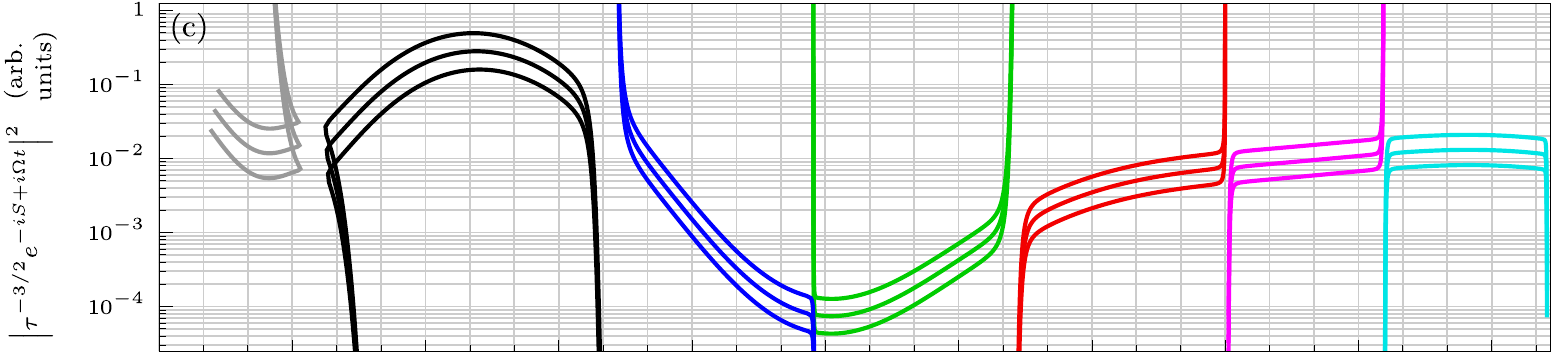}}\\[-2mm]
\subfigure{ \label{fig-single-burst-re-im-tau}
\includegraphics[scale=1]{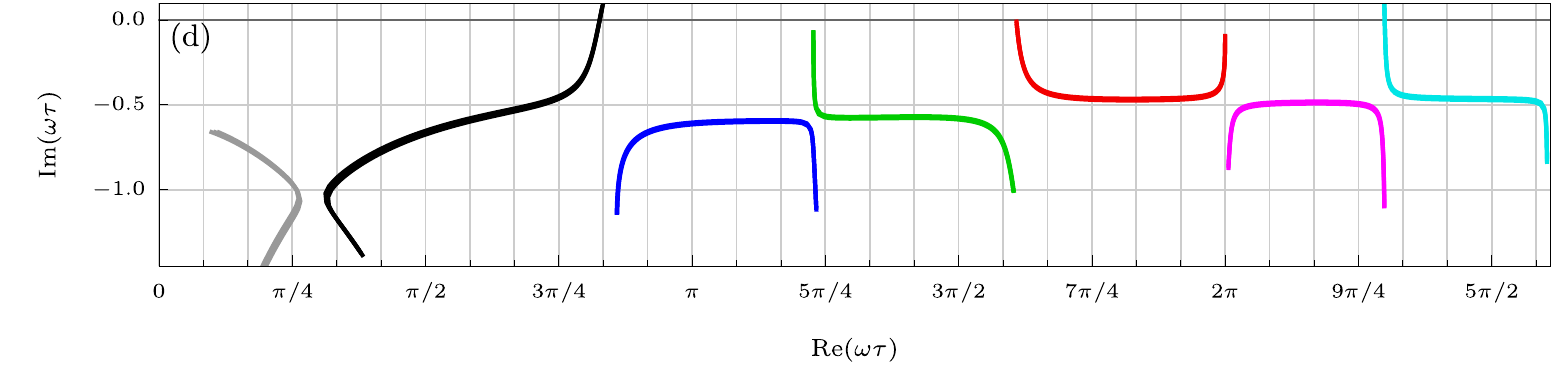}}
\end{tabular}
\caption{
Behaviour of the temporal saddle points in the rotating frame. The harmonic emission as a function of harmonic order (\hyperref[fig-single-burst-intensity-ho]{a}) varies for the multiple contributing quantum orbits; this dependence is best expressed through the (complex) excursion time $\tau=t-t'$, which we plot in (\hyperref[fig-single-burst-re-im-tau]{d}), and which give the well-known energy-time mapping (\hyperref[fig-single-burst-tau-ho]{b}) and the harmonic emission in (\hyperref[fig-single-burst-int-ttau]{c}), which closely resembles the known laboratory-frame results~\cite[Fig.~8]{milosevic-becker-kopold_coplanar-field-mixing_2000}. In (\hyperref[fig-single-burst-intensity-ho]{a}) and (\hyperref[fig-single-burst-int-ttau]{c}) the line triplets correspond to $m=1$, $0$ and $-1$, from top to bottom.
}
\label{fig-intensity-tau}
\end{figure*}

The resulting saddle points closely resemble the quan\-tum-orbit behaviour in the laboratory frame~\cite{milosevic-becker-kopold_coplanar-field-mixing_2000}, and we show the saddle points and their relationship with the harmonic order and the trajectory-determined harmonic emission in \reffig{fig-intensity-tau}. As in the laboratory frame, there are multiple possible quantum orbits, spanning several possible returns of the photoelectron to the ion, but the ionization factor $|e^{-iS}|^2$ strongly selects the shortest quantum orbit (with excursion times $\tau=t-t'$ between $\SI{50}{\degree}/\omega$ and $\SI{150}{\degree}/\omega$, shown in black). The addition of the rotation factor $e^{im\alpha (t-t')}$ to the action shifts the contributions of the $p_\pm$ orbitals by a factor of about $2.3$, which comes from the added ionization potential; there is also a slight shift in the ionization saddle points, but its contribution to the harmonic emission is negligible.

\begin{figure*}[t!]
\centering
\setlength{\tabcolsep}{0.3mm}
\begin{tabular}{cc}
\subfigure{\label{fig-harmonic-spectrum-p-}%
\includegraphics[scale=1]{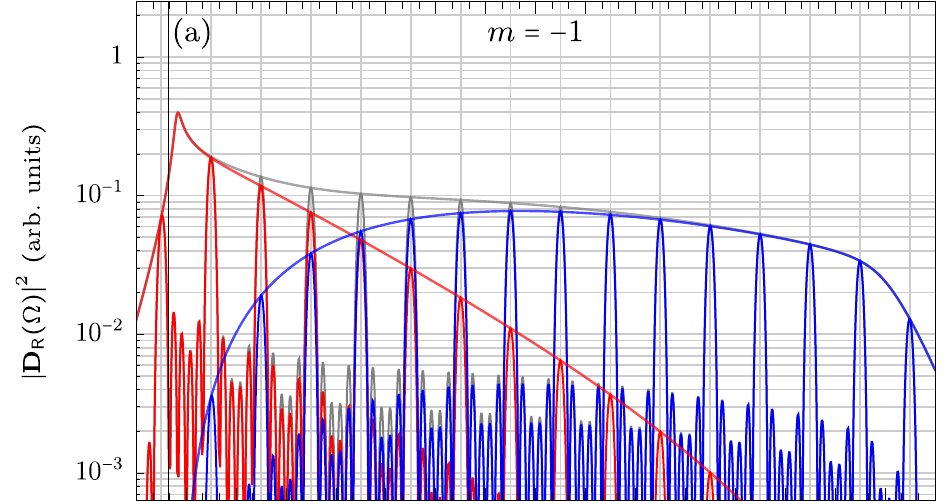}}&
\subfigure{\label{fig-harmonic-spectrum-p+}%
\includegraphics[scale=1]{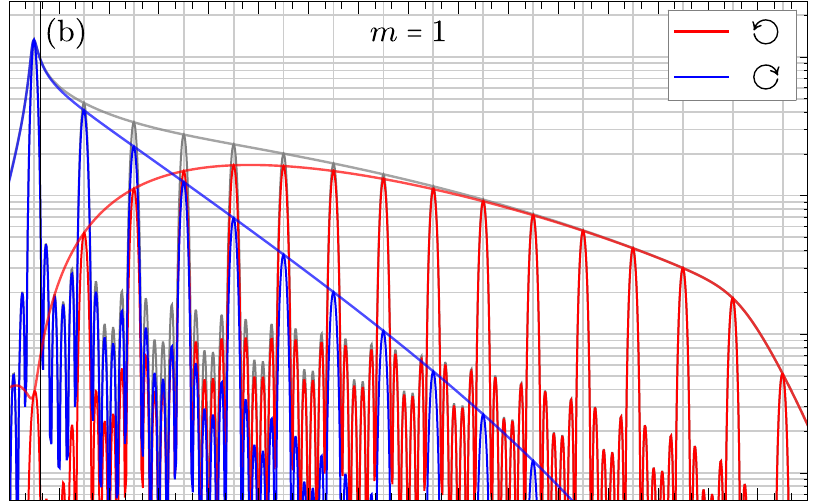}}\\[-1.4mm]
  \includegraphics[scale=1]{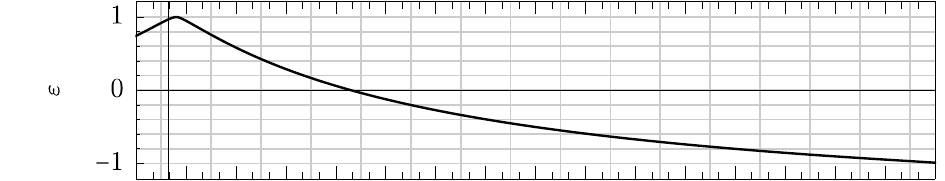}&
  \includegraphics[scale=1]{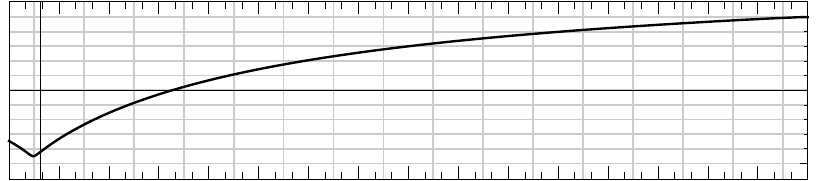}\\[-0.5mm]
\subfigure{\label{fig-harmonic-spectrum-s}%
\includegraphics[scale=1]{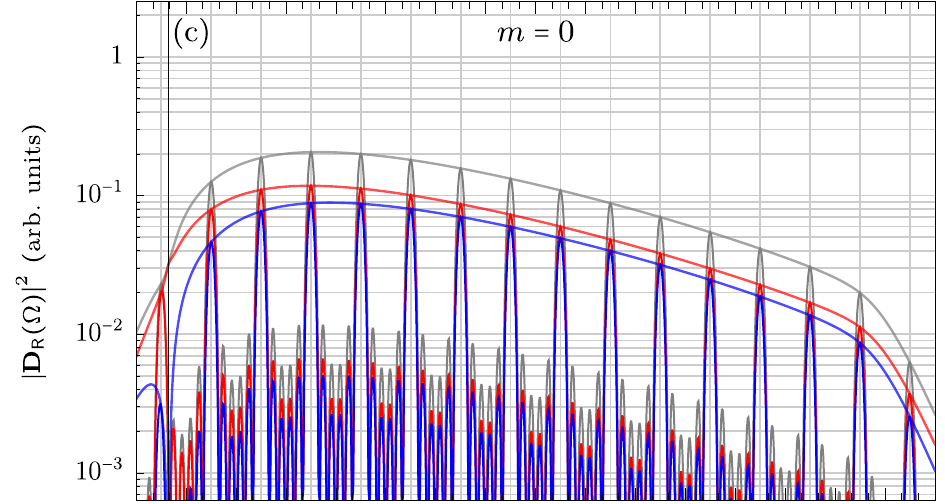}}&
\subfigure{\label{fig-harmonic-spectrum-2p}%
\includegraphics[scale=1]{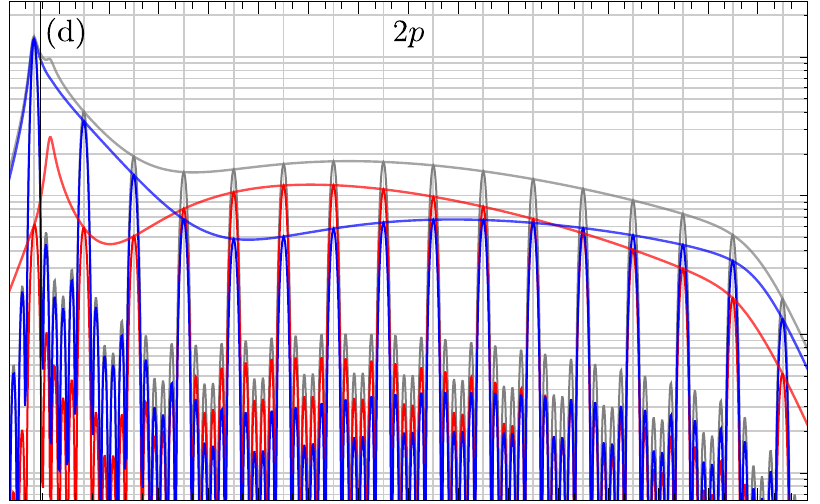}}\\[-1.4mm]
  \includegraphics[scale=1]{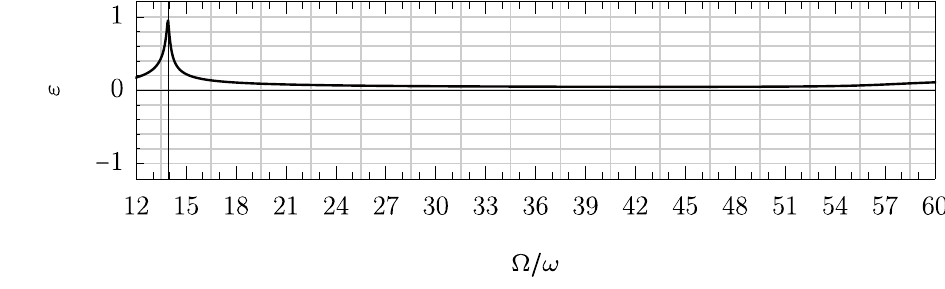}&
  \includegraphics[scale=1]{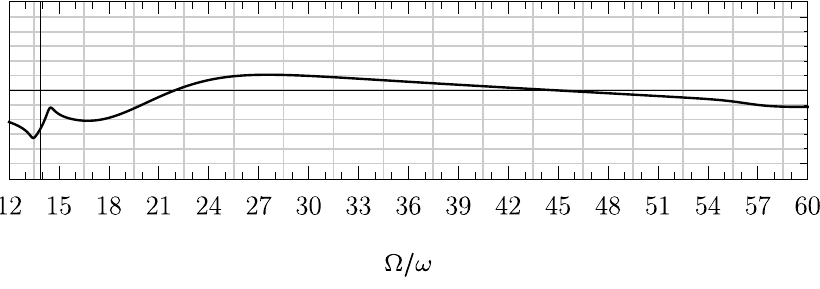}
\end{tabular}
\caption{
Harmonic spectra in the rotating frame for $p_-$, $p_+$ and $s$ orbitals (for neon in a $\SI{800}{nm}$ field of intensity $I=\SI{1.88E14}{W/cm^2}$), and for the coherent addition of both $p$ emissions (a, b, c, and d, resp.), with the right- and left-handed components shown in red and blue (milder and darker gray) respectively, and the total emission in light gray. The harmonics represent two periods of the rotating-frame field (so $6\pi/\omega$ in total) from a monochromatic field, and the continuous lines are the contributions from one single ionization burst. The lower panels show the signed ellipticity of the emission.
}
\label{fig-harmonic-spectra-rotating-frame}
\end{figure*}

Having completed the rotating-frame calculation, to get a concrete spectrum we still require an explicit recombination dipole $\vbd^*\!(\vbk)$ and the ionization matrix element $\Y(\vbk)$ for the $p$ states in question; these we calculate in appendix~\ref{sec-appendix-matrix-elements} for a short-range wavefunction of the form $\braket{ \vbr}{g} = C_{\kappa, \ell}\,Y_{lm}(\hat\vbr) e^{-\kappa r}/\kappa r$, where $\kappa$ is the characteristic momentum of the ionization potential $I_p = \tfrac12 \kappa^2$ and $C_{\kappa,\ell}$ is a normalization factor that is irrelevant for our purposes~\cite{perelomov-ionization-1966}.

\begin{figure*}[t!]
\centering
\setlength{\tabcolsep}{0.3mm}
\begin{tabular}{cc}
\subfigure{\label{fig-polarization-ellipse-2d-p-}%
\includegraphics[scale=1]{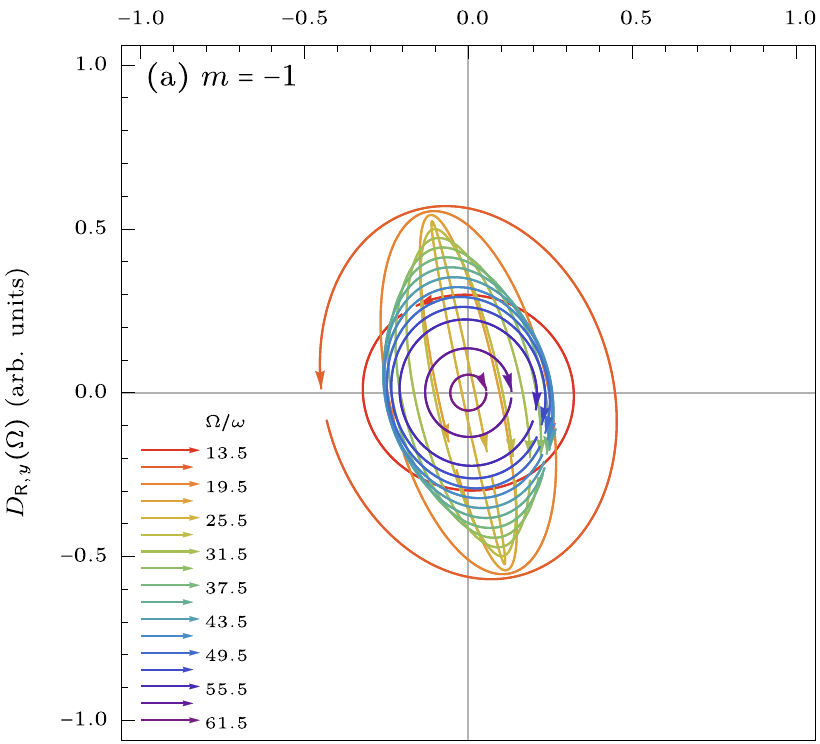}}&
\subfigure{\label{fig-polarization-ellipse-2d-p+}%
\includegraphics[scale=1]{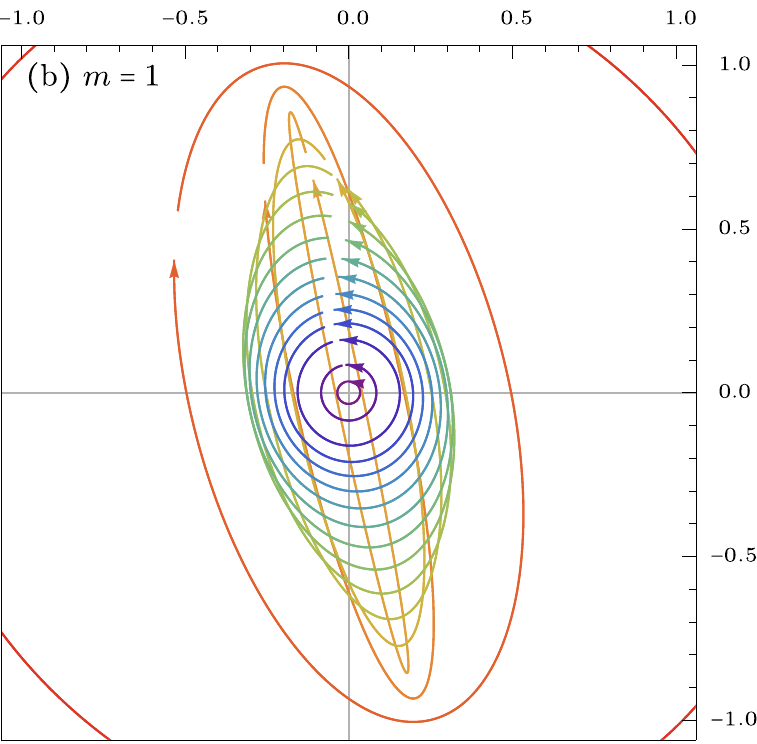}}\\
\subfigure{\label{fig-polarization-ellipse-2d-s}%
\includegraphics[scale=1]{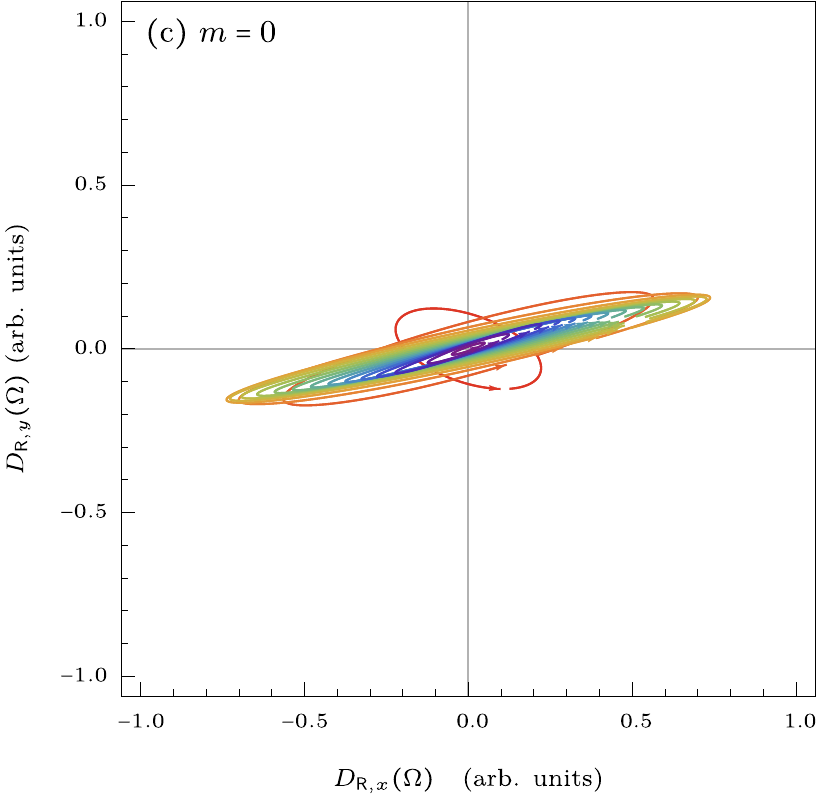}}&
\subfigure{\label{fig-polarization-ellipse-2d-2p}%
\includegraphics[scale=1]{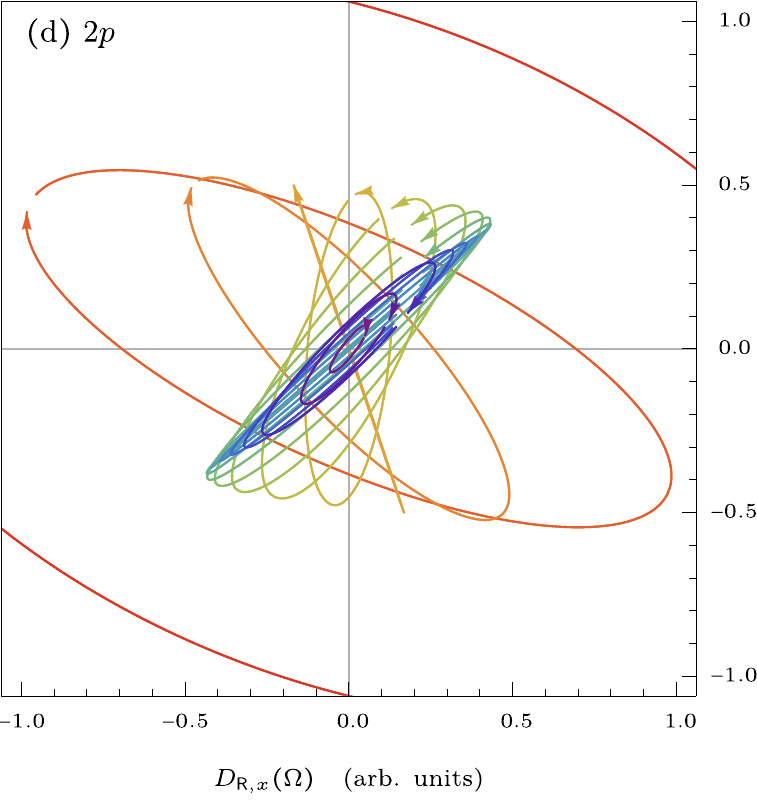}}
\end{tabular}
\caption{
Polarization ellipses of the harmonic emission (defined as the path of $\Re(e^{-i\varphi}\vbD_\rot(\Omega))$ over $\varphi\in[0,2\pi]$), colour-coded over the allowed harmonics in the rotating frame, $\Omega = (2k+1)\frac32 \omega$. The direction of the arrows shows the helicity of the emission; the arrows' position within the ellipse shows $\Re(\vbD_\rot(\Omega))$, excluding the standard atto-chirp phase $e^{-i\Re(S-\Omega t)}$ caused by the quantum-orbit dynamics, so it tracks the phase of the harmonics.
}
\label{fig-polarization-ellipses-2d}
\end{figure*}

\section{Harmonic spectra and polarization}
\label{sec-harmonic-spectra}
Our framework now enables us to calculate the harmonic spectrum in the rotating frame, which we show in \reffig{fig-harmonic-spectra-rotating-frame}, where we consider monochromatic fields of wavelength $\SI{800}{nm}$ and intensity $I=\SI{1.88E14}{W/cm^2}$ acting on neon, with an ionization potential of $\SI{21.6}{eV}$; we also show the emission of a (fictional) $s$ orbital at the same ionization potential, for easier comparison. We examine the contributions of six ionization bursts, spanning one revolution of the rotating frame with respect to the laboratory frame. This harmonic emission closely matches the equivalent laboratory-frame SFA calculation (as shown in \reffig{fig-harmonic-spectra-lab-frame}), as expected, and it is a good match to numerical Schrödinger-equation simulations~\cite{ jimenez_control-of-polarization_2017} (barring a region above the ionization potential at harmonics ${\sim}$13 to 22, where the SFA is known to be unreliable due to its treatment of the continuum as flat plane waves).

The harmonic spectra in the rotating frame quickly show several of the relevant features. For each initial orbital $p_m$, the co-rotating harmonic emission, along $\ue m$, dominates the plateau, while the counter-rotating emission along $\ue{-m}$ drops on a steep exponential, after dominating the harmonic emission at threshold ($\Omega\gtrsim I_p$) and the early plateau, where the SFA is unreliable. In the mid-plateau, the right-handed emission from the $p_+$ orbital dominates, giving an overall right-handed spectrum, but its contribution drops slightly faster than the $p_-$ emission, which dominates closer to the harmonic cut-off. %
This effect can also be seen on numerical simulations~\cite{zhang_helicity-reversion_2016, jimenez_control-of-polarization_2017} and its appearance here in calculations with only the short quantum orbit (with the second return producing harmonics over one order of magnitude weaker, as seen in \reffig{fig-single-burst-int-ttau}) provides an alternative to the existing explanations based on the second return's slightly higher harmonic cutoff at excursion times around $\SI{300}{\degree}/\omega$~\cite{zhang_helicity-reversion_2016}.

\begin{figure*}[t!]
\centering
\setlength{\tabcolsep}{2mm}
\begin{tabular}{cc}
\subfigure{\label{fig-polarization-ellipse-3d-p-}%
\includegraphics[scale=1]{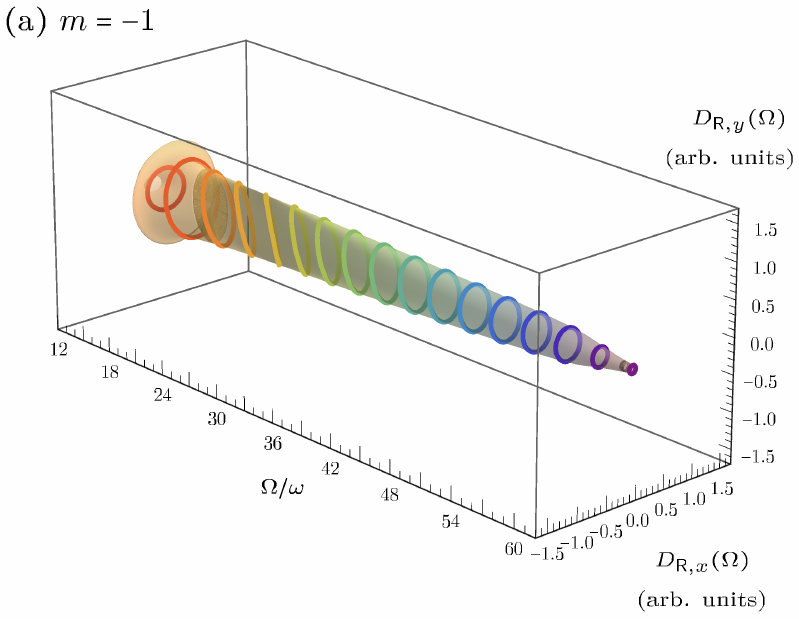}}&
\subfigure{\label{fig-polarization-ellipse-3d-p+}%
\includegraphics[scale=1]{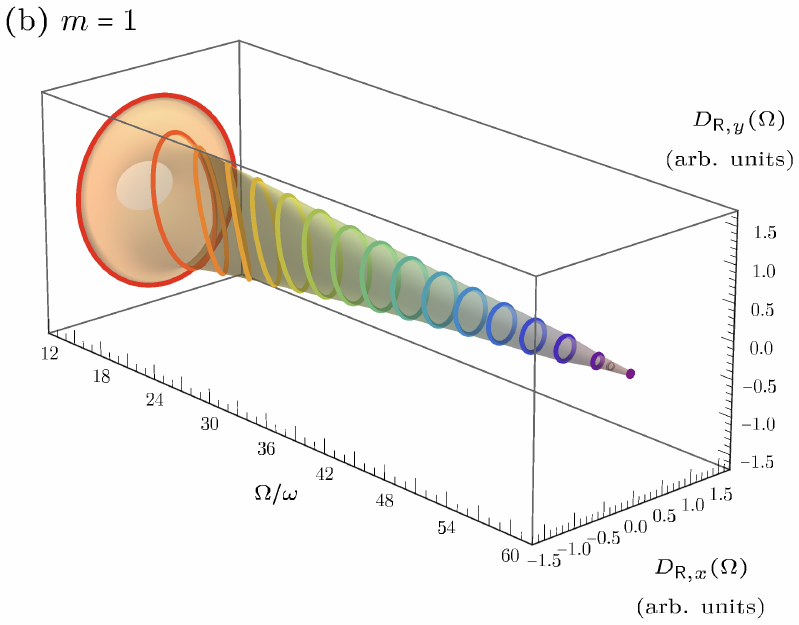}}\\[-1mm]
\subfigure{\label{fig-polarization-ellipse-3d-s}%
\includegraphics[scale=1]{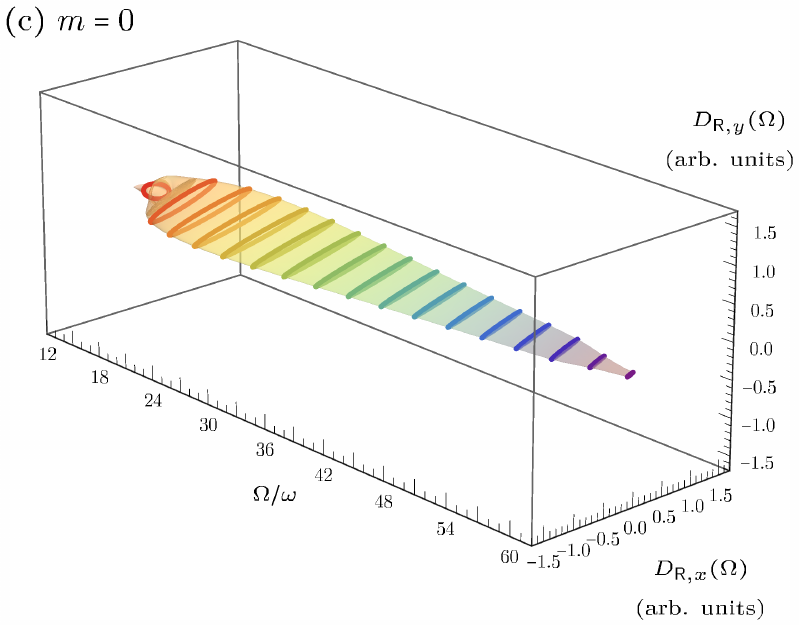}}&
\subfigure{\label{fig-polarization-ellipse-3d-2p}%
\includegraphics[scale=1]{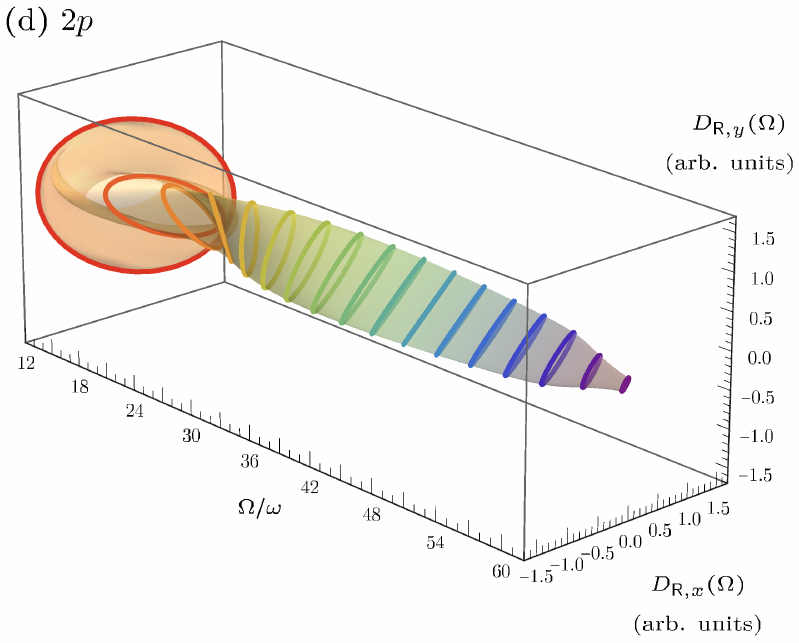}}
\end{tabular}
\caption{
Polarization ellipses of the harmonic emission, as shown in \reffig{fig-polarization-ellipses-2d}, broken out over the harmonic order to avoid overlaps.
}
\label{fig-polarization-ellipses-3d}
\end{figure*}

In addition to the harmonic spectra, however, the rotating frame also affords us a more powerful tool to study the harmonic emission -- the polarization of the different lines. In the laboratory frame, the two elements of each pair appear separately, but the transformation to the rotating frame shifts them by $\pm \alpha$ so that they overlap at odd multiples of $\frac32\omega$, allowing us to study the helicity asymmetry of the pair as simply the ellipticity of each rotating-frame line, which we show as the lower panels of \reffig{fig-harmonic-spectra-rotating-frame}. Thus we see that, despite a mid-plateau dominance of the right-polarized emission, the ellipticity of the global harmonic emission never exceeds $\varepsilon\approx 0.2$.

More interestingly, we can also examine the polarization ellipses directly, which we show in Figs.~\ref{fig-polarization-ellipses-2d} and~\ref{fig-polarization-ellipses-3d}, and which exhibit several unexpected features. Most immediately, the spectra from \reffig{fig-harmonic-spectra-rotating-frame}, which coincide with features known from the laboratory frame, require the existence of a linearly polarized line for both $p_\pm$ emissions, at the point where the right- and left-handed amplitudes cross (harmonics $\sim$25 for $p_-$ and $\sim$22 for $p_+$, respectively), and intuition would suggest, given that in the rotating frame the system is driven by an electric field along the $x$ axis, that this linearly-polarized emission would follow that direction. However, the observed emission is orthogonal to that, rotated by a few degrees off of the $y$ axis for both $p$-state emissions.

The global $2p$ emission also shows unexpected features, in the form of a consistent rotation of the polarization ellipse throughout the plateau, which arises from the relative phase of the $\ue{\pm}$-polarized contributions of the two $p_\pm$ orbitals, and which changes across the harmonic emission. This effect has so far gone unnoticed, but it should be measurable through interferometric measurements of the harmonic phase of the different members of the line doublets~\cite{chen-tomographic-2016}.

On the other hand, the $s$-state emission shown in \reffig{fig-polarization-ellipse-2d-s}  is largely linearly polarized along the rotating-frame electric field, with a rotation by a few degrees which can be attributed to the effects of the Coriolis force. More notably, this polarization study clearly shows that the $s$-state harmonic emission is elliptically polarized, which reflects the slight helicity asymmetry observed in experiments~\cite{fleischer_spin_2014, fan_bright-circularly_2015, baykusheva_bicircular-hhg-spectroscopy} and numerical simulations~\cite{baykusheva_bicircular-hhg-spectroscopy, medisauskas_generating_2016, jimenez_control-of-polarization_2017}.

For the $p$-state emissions, the polarization is fixed by the recombination transition dipole moment $\vbd^*(\vbk)$, which is itself given, in Eq.~\eqref{p-state-transition-dipole-final} in terms of the circular polarization vectors~$\ue\pm$ with different weights, so an elliptical polarization is not surprising. For the $s$-state emission, on the other hand, the recombination dipole is fixed by the rotational symmetry of the ground state, which requires it to lie along the recombination velocity $\vbk_\mathrm{r}$, as per~Eq.~\eqref{s-state-transition-dipole-final}:
\begin{align}
\vbd^*\!(\vbk_\mathrm{r}) 
\propto
\frac{
  \vbk_\mathrm{r}
  }{
  (k_\mathrm{r}^2+\kappa^2)^2
  }
.
\end{align}
This means that, if the recombination velocity $\vbk_\mathrm{r}$ were real-valued (or even a complex multiple of a real-valued vector), then its components along $\ue+$ and $\ue-$ would have equal magnitudes, the harmonic emission would be linearly polarized in the rotating frame, and there would be no helicity asymmetry for this case. Since this is in contradiction to the results, we conclude that the recombination velocity must have a nonzero imaginary component that is linearly independentto its real part,$\!$%
\footnote{In fact, the real and imaginary parts of the recombination velocity $\vbk_\mathrm{r}$ must be orthogonal, since the recombination saddle-point equation requires that $\frac12 \vbk_\mathrm{r}^2 = \frac12\left[\Re(\vbk_\mathrm{r})^2-\Im(\vbk_\mathrm{r})^2\right]+i\Re(\vbk_\mathrm{r})\cdot\Im(\vbk_\mathrm{r}) = \Omega-I_p$ be real.}
 and that the $s$-state helicity asymmetry is a direct witness of this fact.

\begin{figure*}[htbp]
\centering
\setlength{\tabcolsep}{-0.2mm}
\begin{tabular}{ccc}
\subfigure{ \label{fig-recombination-dipole-p-}
  \includegraphics[scale=1]{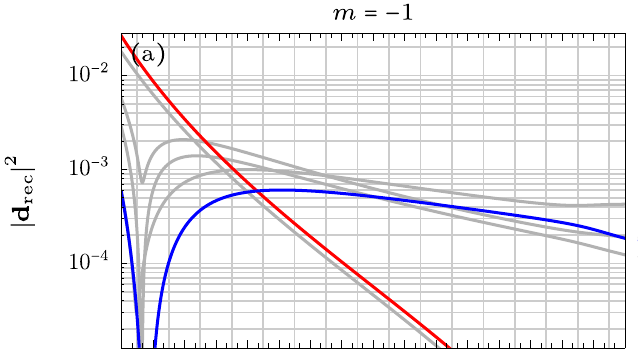}}&
\subfigure{ \label{fig-recombination-dipole-s}
  \includegraphics[scale=1]{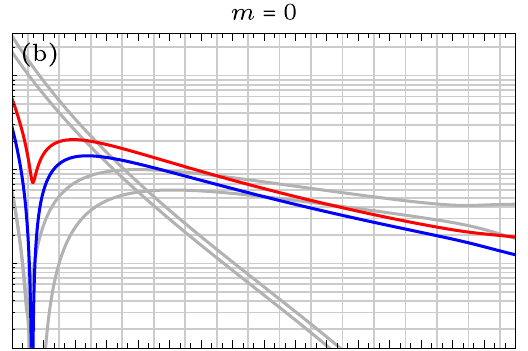}}&
\subfigure{ \label{fig-recombination-dipole-p+}
  \includegraphics[scale=1]{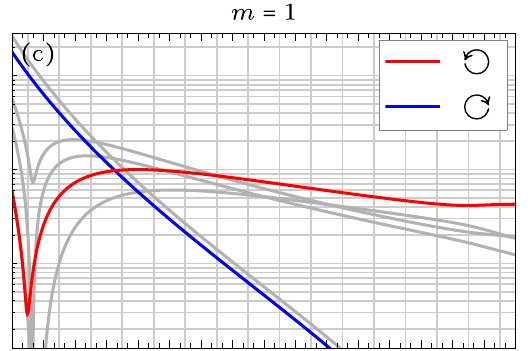}}\\[-3mm]
\subfigure{ \label{fig-ionization-factor-p-}
  \includegraphics[scale=1]{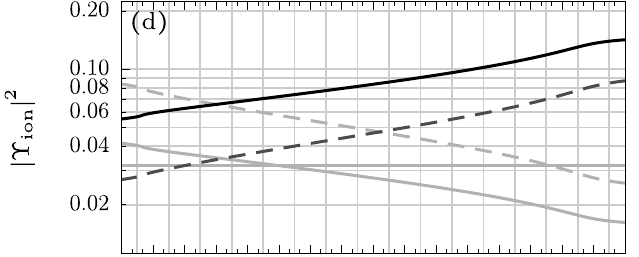}}&
\subfigure{ \label{fig-ionization-factor-s}
  \includegraphics[scale=1]{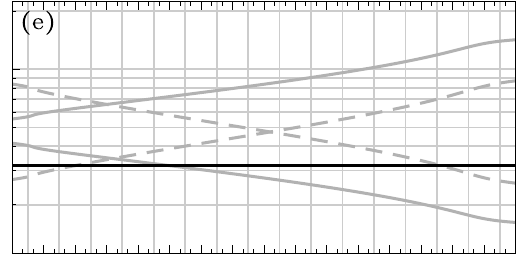}}&
\subfigure{ \label{fig-ionization-factor-p+}
  \includegraphics[scale=1]{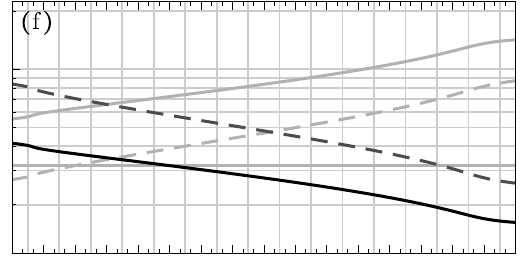}}\\[-3mm]
\subfigure{ \label{fig-action-factor-p-}
  \includegraphics[scale=1]{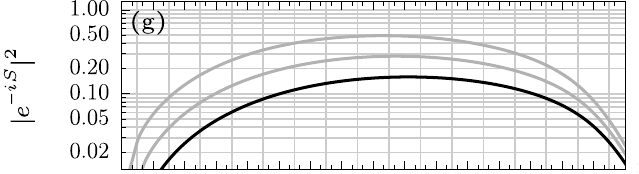}}&
\subfigure{ \label{fig-action-factor-s}
  \includegraphics[scale=1]{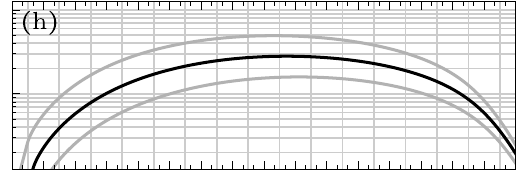}}&
\subfigure{ \label{fig-action-factor-p+}
  \includegraphics[scale=1]{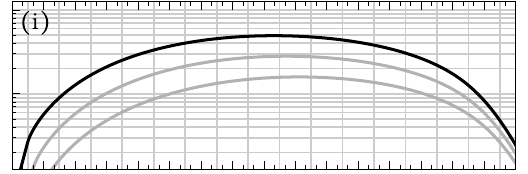}}\\[-3mm]
\subfigure{ \label{fig-total-factor-p-}
  \includegraphics[scale=1]{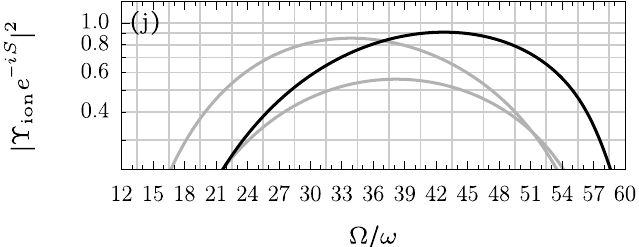}}&
\subfigure{ \label{fig-total-factor-s}
  \includegraphics[scale=1]{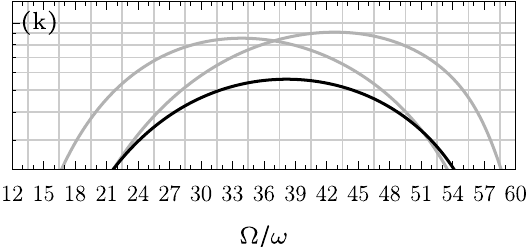}}&
\subfigure{ \label{fig-total-factor-p+}
  \includegraphics[scale=1]{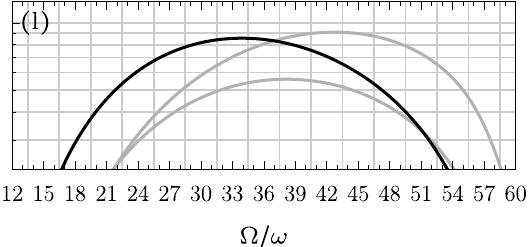}}
\end{tabular}
\caption{
Geometrical factors affecting the helicity asymmetry and the harmonic intensity for the different initial ground-state orbitals, in atomic units. %
(\hyperref[fig-recombination-dipole-p-]{a}-\hyperref[fig-recombination-dipole-p+]{c}) %
The recombination dipole $|\vbd_\mathrm{rec}|^2=|\vbd_{lm}^*   \mathopen{}\left( R_z^{-1}(\alpha (t-t'))\vbp_{\rot,s}(t,t')+\vba_\rot(t))  \right)\mathclose{}|^2$. %
(\hyperref[fig-ionization-factor-p-]{d}-\hyperref[fig-ionization-factor-p+]{f}) %
The ground-state ionization factor $|\Y_\mathrm{ion}|^2 = |\Y_{lm}\mathopen{}\left(  \vbp_{\rot,s}(t,t')+\vba_\rot(t'))\right)\mathclose{}|^2$. The solid black lines indicate the ionization factor while the dashed black lines show the ionization factors with an exponential term removed, $|e^{-im\alpha t'}\Y_\mathrm{ion}|^2$, which closely match the laboratory-frame versions of Figs.~\ref{fig-internal-structure-of-spectra-lab-frame}\subref{fig-lab-frame-ionization-factor-p-} and~\subref{fig-lab-frame-ionization-factor-p+}. %
(\hyperref[fig-action-factor-p-]{g}-\hyperref[fig-action-factor-p+]{i}) %
Complex exponential of the action, in arbitrary units. %
(\hyperref[fig-total-factor-p-]{j}-\hyperref[fig-total-factor-p+]{l}) Complex exponential of the action multiplied by the ground-state ionization factor, in arbitrary units. In all panels, the grey lines are plotted for visual reference and indicate the lines of the other columns.}
\looseness=-1
\label{fig-internal-structure-of-spectra}
\end{figure*}

\section{Helicity asymmetry from the quantum-orbit dynamics}
\label{discussion}
Having explored the main features of the harmonic spectra in the rotating frame, we now turn to their origins within the SFA expression for the harmonic dipole, and what they tell us about the rotating-frame harmonic emission.

The most immediate feature of the harmonic spectra shown in \reffig{fig-harmonic-spectra-rotating-frame} is the swift drop-off of the counter-rotating harmonics polarized as $\ue\mp$ in the $p_\pm$ emission. This can only be caused by the recombination dipole (since it is the only part of the SFA expression for $\vbD_\rot(\Omega)$ that affects the polarization), and we show its behaviour in Figs.~\ref{fig-internal-structure-of-spectra}(\hyperref[fig-recombination-dipole-p-]{a}-\hyperref[fig-recombination-dipole-p+]{c}). This swift decay is a consequence of the different strengths of the Clebsch-Gordan coefficients for different angular momenta, which favors the  emission of a photon with an angular momenta rotating in the same direction as the target state~\cite{jimenez_control-of-polarization_2017}. It is also influenced by the interference between the two $\ue m^*$ components in~\eqref{p-state-transition-dipole-final}~-- the $s$-wave component in $S_{00}$ and the $d$-wave component in $S_{20}$ -- and as such it is relatively fragile to effects coming from scattering phases in more structured continua, especially for low energies. We also note that in the rotating frame the counter-rotating lines for $p_+$ and $p_-$ differ by $\omega$ due to the change in the $I_p$, in contrast to the laboratory frame where they are equal.

In addition to this, there is a clear dominance of the right-handed recombination dipole from the $p_+$ emission (red (milder gray) in \reffig{fig-recombination-dipole-p+}) over the left-handed recombination dipole from the $p_-$ emission (blue (darker gray) in \reffig{fig-recombination-dipole-p-}), which is modest but sustained throughout the range of the emission. To understand the origin of this imbalance, we look at the recombination dipoles from~\eqref{p-state-transition-dipole-final}, which tells us that those dominating components are given, up to common factors,~by
\begin{align}
\ue{m}^*\cdot\vbd_{1m}^*(\vbk_\mathrm{r}) 
& \propto
S_{2,-2m}(\vbk_\mathrm{r})
\propto
(k_{\mathrm{r},x} -m i k_{\mathrm{r},y})^2
\nonumber \\ & 
\propto
\left[\ue{m}^*\cdot \vbk_{\mathrm{r}}\right]^2
=
k_{\mathrm{r},\pm}^2
,
\end{align}
where $\vbk_\mathrm{r}$ is the recollision velocity. The component $k_{\mathrm{r},\pm} =\ue{\pm}^*\cdot \vbk_{\mathrm{r}}=\frac{\mp 1}{\sqrt{2}}(k_{\mathrm{r},x} \mp i k_{\mathrm{r},y})$ of the recollision velocity is not an intuitive object, but we have already encountered it, through the elliptical polarization of the $s$-state emission, where it diagnosed a nonzero imaginary part of the recollision velocity as responsible for the imbalance in the $s$-state recollision dipole shown in \reffig{fig-recombination-dipole-s}. For the dominant $\ue\pm$ component of $p_\pm$-state emission, this factor is squared, giving twice the separation, and this is again caused by the fact that the recollision velocity is complex-valued~\cite{jimenez_control-of-polarization_2017}.

However, for $p$ states, this is not the end of the story, because the dominant emission on each circular component comes from a different orbital, and the two $p$ ground states ionize at different rates, shown in Figs.~\ref{fig-internal-structure-of-spectra}(\hyperref[fig-total-factor-p-]{j}-\hyperref[fig-total-factor-p+]{l}). In terms of the harmonic spectrum, this can be seen in the dominance of left-polarized harmonics near the cutoff in the total harmonic emission of \reffig{fig-harmonic-spectrum-2p}, despite the fact that the $p_+$ recombination dipole dominates throughout the spectrum.

To understand the ionization rate that causes these differences, we separate it into its two natural factors: the regularized matrix element $\Y(\vbk_\mathrm{i}) = (\frac12\vbk_\mathrm{i}^2+I_p)\braket{\vbk_\mathrm{i}}{g}$, where $\vbk_\mathrm{i}$ is the ionization velocity, and the tunnelling action $e^{-iS}$, which is affected by the $m$-dependent ionization potential. We have already explored the action in \reffig{fig-single-burst-intensity-ho}, and we recapitulate its behaviour over the short quantum orbits in Figs.~\ref{fig-internal-structure-of-spectra}(\hyperref[fig-action-factor-p-]{g}-\hyperref[fig-action-factor-p+]{i}).

In addition to this global shift in the action, however, there is also a strong dependence in the reduced matrix element, as shown in Figs.~\ref{fig-internal-structure-of-spectra}(\hyperref[fig-ionization-factor-p-]{d}-\hyperref[fig-ionization-factor-p+]{f}): this includes both an overall prevalence of the $p_-$ ionization, opposite to the suppression from the action, as well as a rolling dependence on the harmonic order, with the $p_+$ becoming even more suppressed towards the cutoff. This rolling dependence is very similar to the laboratory-frame dependence of the ionization factor $\Y(\vbk_\mathrm{i})$, with an offset -- and, in fact, if we factor out the exponential factor of $e^{-im\alpha t'}$ from the action and we add it to the $\Y(\vbk_\mathrm{i})$, as shown dashed in Figs.~\ref{fig-internal-structure-of-spectra}\subref{fig-ionization-factor-p-} and~\subref{fig-ionization-factor-p+}, the match to the laboratory-frame ionization factor of Figs.~\ref{fig-internal-structure-of-spectra-lab-frame}(\hyperref[fig-lab-frame-ionization-factor-p-]{d}-\hyperref[fig-lab-frame-ionization-factor-p+]{f}) is essentially exact. As such, the total ionization amplitude $|\Y e^{-iS}|^2$, shown in Figs.~\ref{fig-internal-structure-of-spectra}(\hyperref[fig-total-factor-p-]{j}-\hyperref[fig-total-factor-p+]{l}), does not change when moving to the rotating frame.

This tells us, then, that in addition to affecting the ionization potential through an effective magnetic field, the transformation to the rotating frame also has a strong effect on the ionization matrix element $\Y(\vbk_\mathrm{i}) \propto\braket{\vbk_\mathrm{i}}{g}$ -- and, moreover, that this effect exactly cancels out that of the effective magnetic field.

In essence, the change in the ionization matrix element is caused by the rotation of the ionization velocity $\vbk_{\lab,\mathrm{i}}$ in the laboratory frame to the rotating frame, via the rotation
\begin{equation}
\vbk_{\rot,\mathrm{i}} = R_z^{-1}(\alpha t')\vbk_{\lab,\mathrm{i}},
\end{equation}
except that now, because the ionization time $t'$ is complex, the frame transformation must now go over a complex angle $\alpha t'$. For real angles, the eigenvectors of the rotation $R_z^{-1}(\alpha t')$ are the circular unit vectors $\ue\pm$, but this relationship holds for all complex-valued rotation angles, because the trigonometric algebra that underpins the eigenvalue relation in Eq.~\eqref{rotation-of-unit-circular-vectors}, will work for any arbitrary complex $\alpha t'$. However, if the rotation angle $\alpha t'$ is imaginary, the eigenvalues $e^{\mp i \alpha t'}$ are no longer pure phases: instead, they become amplitude factors that affect exponentially the size of the component along each circular unit vector after the transformation. Thus, for positive $\Im(t')$, the $\ue+$ component is exponentially enhanced, while the $\ue-$ component is suppressed.

To understand these changes, it is helpful to look at the explicit frame transformation when the rotation angle is large and positive-imaginary, in which case it takes the form
\begin{align}
R_z^{-1}(i\alpha \tau) 
& =
\begin{pmatrix}
 \cosh(\alpha \tau) & -i\sinh(\alpha \tau) & 0 \\
 i\sinh(\alpha \tau) & \phantom{i}\cosh(\alpha \tau) & 0 \\
0 & 0 & 1
\end{pmatrix}
\nonumber \\ & \approx
\frac{1}{2}
e^{\alpha \tau}
\begin{pmatrix}
 1 & -i & 0 \\
 i &  1 & 0 \\
0 & 0 & 0
\end{pmatrix}
.
\end{align}
Here the positive exponential factor overwhelms the rest of the matrix, and it is left multiplying the projector $\ue+^{\phantom{\dagger}}\ue+^{\dagger}$, which turns any real-valued vector into a multiple of the right-handed unit vector $\ue+$. Geometrically speaking, the action of this transformation on any real vector amplifies it and gives it an imaginary component $\SI{90}{\degree}$ counter-clockwise from its real part; the same is true (approximately) for the full hyperbolic-functions matrix on the left. %
(Similarly, the action of $R_z^{-1}(-i\alpha \tau)$ on a real vector gives it an imaginary component directed clockwise from its real part.)

\begin{figure*}[htbp]
\centering
\setlength{\tabcolsep}{0.1mm}
\begin{tabular}{rl}
\subfigure{ \label{fig-trajectory-position-17}
  \includegraphics[scale=1]{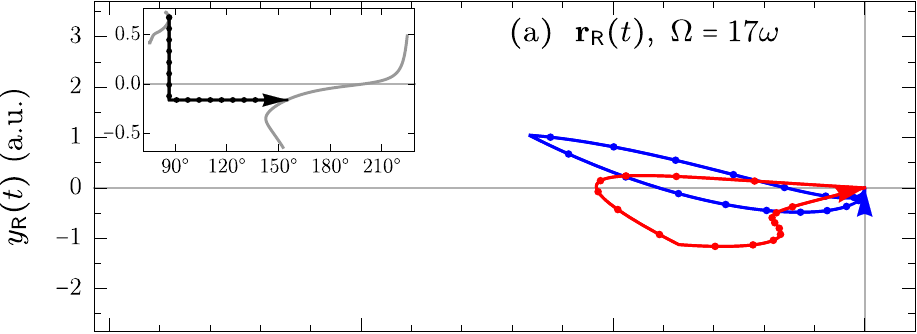}}&
\subfigure{ \label{fig-trajectory-velocity-17}
  \includegraphics[scale=1]{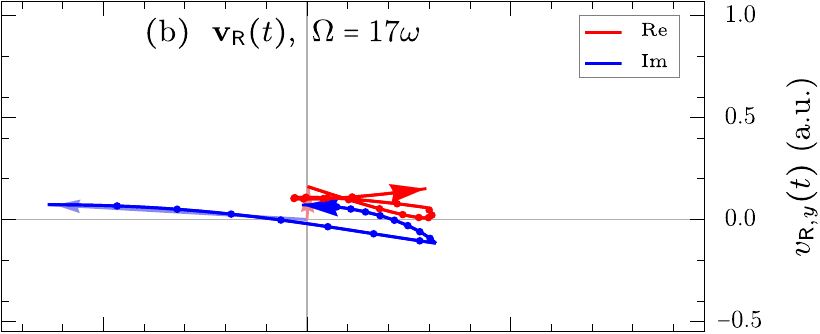}}\\[-2.7mm]
\subfigure{ \label{fig-trajectory-position-27}
  \includegraphics[scale=1]{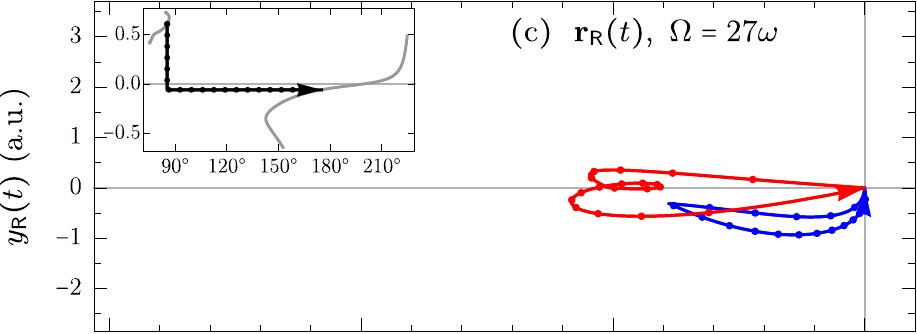}}&
\subfigure{ \label{fig-trajectory-velocity-27}
  \includegraphics[scale=1]{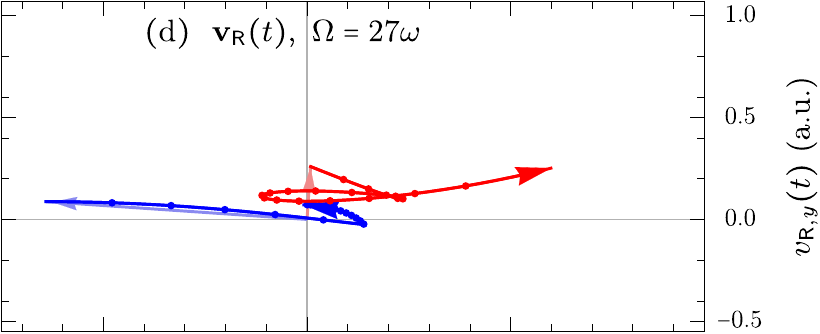}}\\[-2.7mm]
\subfigure{ \label{fig-trajectory-position-37}
  \includegraphics[scale=1]{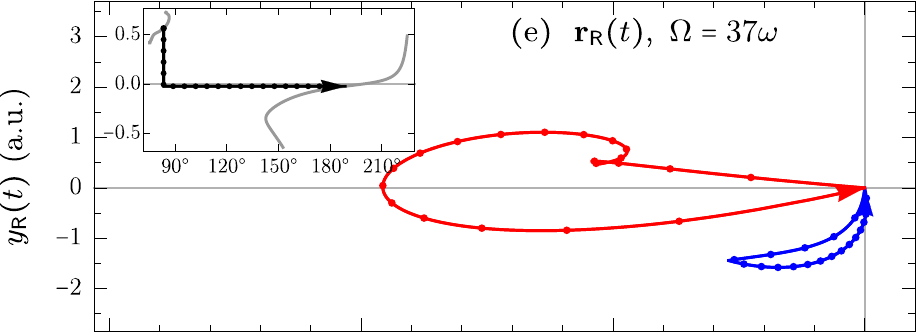}}&
\subfigure{ \label{fig-trajectory-velocity-37}
  \includegraphics[scale=1]{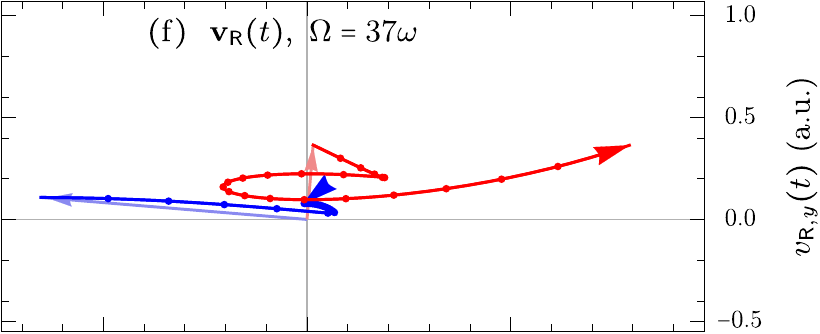}}\\[-2.7mm]
\subfigure{ \label{fig-trajectory-position-47}
  \includegraphics[scale=1]{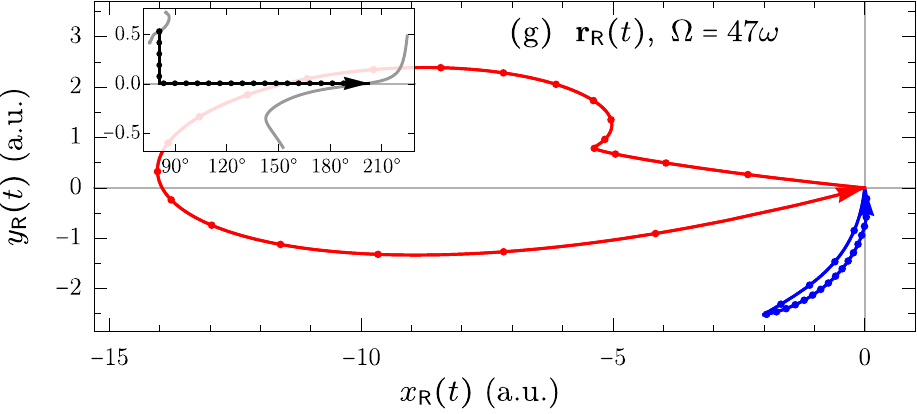}}&
\subfigure{ \label{fig-trajectory-velocity-47}
  \includegraphics[scale=1]{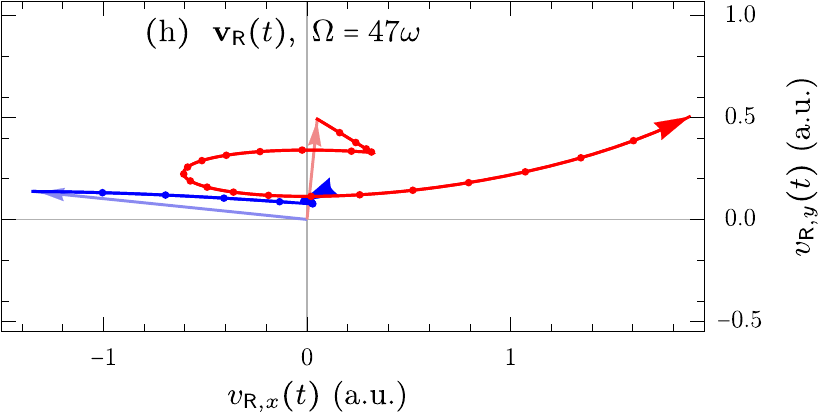}}
\end{tabular}
\caption{
Photoelectron trajectories in the rotating frame, corresponding to harmonic orders 17,~27,~37 and 47, both over position and velocity (left and right columns, resp.), in atomic units. We show the real and imaginary parts of the trajectory in red and blue (milder and darker gray), respectively, and we take the trajectory over the standard contour shown in the inset, between the ionization and recombination times of the short quantum orbit (shown in gray); the equispaced dots mark progress along the contour. In the right-hand column, the pale arrows mark the start of the trajectory, i.e. the ionization velocity.
}
\vspace{-5mm}
\label{fig-trajectories}
\end{figure*}

More importantly, the ionization matrix element is directly proportional to the circular components of $\vbk$, since the solid harmonic involved,
\begin{equation}
\Y_{1m}(\vbk_\mathrm{i})
\propto
S_{1m}(\vbk_\mathrm{i})
\propto
\ 
k_{\mathrm{i},x} +m i k_{\mathrm{i},y}
,
\end{equation}
is nothing more than the rotating-frame ionization velocity's component along the circular unit vector $\ue{-m}$, as shown in Eq.~\ref{r-in-circular-basis}. As discussed above, the eigenvalue associated to the eigenvector $\ue{-m}$ of the rotation is the factor $e^{+i m \alpha t'}$, exactly the opposite to that induced in the action by the ionization-potential change.

These considerations can be brought to the fore more clearly by examining the quantum-orbit complex trajectories
\begin{align}
\vbr_\rot(t) 
& = R_z^{-1}(\alpha t)\vbr_\lab(t)
= R_z^{-1}(\alpha t) \int_{t_\mathrm{i}}^{t} \left[\vbp_\lab + \vba(\tau)\right]\d\tau
\nonumber \\ & = R_z^{-1}(\alpha (t-t_\mathrm{i})) \int_{t_\mathrm{i}}^{t} \big[\vbp_\rot + R_z(\alpha (\tau-t_\mathrm{i}))  \vba_\rot(\tau)\big]\d\tau
\end{align}
responsible for the harmonic generation, which we show in \reffig{fig-trajectories}, and whose real parts resemble the recolliding quantum orbits in an elliptical field~\cite{salieres_feynman_2001}. As far as the circular components $k_{\mathrm{i},\pm}$ are concerned, though, the most important aspect is the chiral interplay between the real and imaginary parts of the trajectory and its velocity and, specifically, whether the rotation from the real to the imaginary part is mostly clockwise or counter-clockwise.

In the rotating-frame quantum orbits, this chiral relation is clear and constant: the imaginary part of the position, shown in blue (darker gray), goes off towards the negative $y$ direction, counterclockwise from the real part. In contrast, this effect does not appear in the laboratory frame, shown in \reffig{fig-trajectories-lab-frame}, where the real and imaginary parts of the trajectory can and do lie on either side of each other.

This behaviour is also visible when we examine the velocity, in Figs.~\ref{fig-trajectories}(\hyperref[fig-trajectory-velocity-17]{b},\,\hyperref[fig-trajectory-velocity-27]{d},\,\hyperref[fig-trajectory-velocity-37]{f},\,\hyperref[fig-trajectory-velocity-47]{h}): the real and imaginary parts of the ionization velocity (shown as pale red and blue arrows, respectively) are also in a clear chiral relation, with the imaginary part $\SI{90}{\degree}$ counterclockwise from the real part: this is the fundamental chiral asymmetry which produces the enhanced $\Y(\vbk_\mathrm{i})$ ionization factor for the $p_-$ orbital in the rotating frame, and which is absent in the la\-bo\-ra\-to\-ry-frame trajectories of Figs.~\ref{fig-trajectories-lab-frame}(\hyperref[fig-lab-frame-trajectory-velocity-17]{b},\,\hyperref[fig-lab-frame-trajectory-velocity-27]{d},\,\hyperref[fig-lab-frame-trajectory-velocity-37]{f},\,\hyperref[fig-lab-frame-trajectory-velocity-47]{h}).

It is also important to note that this chiral relationship is independent of the choice of contour that joins the complex ionization and recollision times, which is in principle arbitrary. Measures based on the sense of rotation of the trajectories are somewhat fragile in this regard, but a change in the contour will multiply $\vbk_\mathrm{i}$ and $\vbk_\mathrm{r}$ by a complex number, which will not alter the chiral relationship between their real and imaginary parts.

More generally, it is important to remark that the traditional SFA splitting of the harmonic dipole into ionization, propagation and recombination factors~\cite{HHGTutorial} changes when we move from the usual laboratory frame, as in \reffig{fig-internal-structure-of-spectra-lab-frame}, to the rotating frame, as in \reffig{fig-internal-structure-of-spectra}. That is, these changes tell us that this factorization, being dependent on the frame of reference, is artificial, and does not carry strict physical meaning.

Similarly, the recollision velocity also exhibits a persistent chiral asymmetry throughout the harmonic spectrum: in a sense this is weaker, since the imaginary part of the recollision velocity is smaller, but it is also more robust, because the recollision time is largely real and this means that the behaviour remains in the laboratory frame. This is the fundamental chiral asymmetry that is responsible for the ellipticity of the $s$-state in the rotating frame, and therefore also for the helicity asymmetry in the harmonic emission of helium in bicircular fields.

To summarize, then, the SFA formalism can be cleanly re-expressed in the rotating frame to bring fresh insights into the harmonic emission in bicircular fields. The added Coriolis term shifts the contributions of the different orbitals, but this effect is exactly canceled out by the complex-angle rotation of the ionization velocity, which introduces exponential changes to the amplitude of its two circular components, with strong implications for the quantum-orbits theory of ionization in circularly-polarized fields. Furthermore, the joining of the line doublets in the rotating frame enables us to perform a polarimetric analysis to get additional insights -- notably, that the $s$-state helicity asymmetry is directly caused by the imaginary part of the recollision velocity -- as well as obtain observations -- like the rotation of the polarization axis of the full $2p$ emission -- that are amenable to experimental testing.

\subsection*{Acknowledgements}
We thank Daniel M. Reich and Misha Ivanov for helpful conversations. %
%
EP acknowledges support from %
MINECO grants 
FISICATEAMO (FIS2016-79508-P) and 
Se\-ve\-ro Ochoa (SEV-2015-0522), %
Fundació Cellex, %
Ge\-ne\-ra\-li\-tat de Ca\-ta\-lu\-nya (2014 SGR 874 and CERCA/ Program), %
and ERC grants %
EQuaM (FP7-ICT-2013-C No.~323714), 
QUIC (H2020-FET\-PRO\-ACT{-}2014 No. 641122) %
and %
OSYRIS (ERC-2013-ADG No.~339106)%
.
AJG acknowledges funding from DFG QUTIF grant IV 152/6-1. %

\appendix

\renewcommand{\theequation}{\Alph{section}.\arabic{equation}}
\numberwithin{equation}{section}

\numberwithin{figure}{section}
\renewcommand{\thesubfigure}{(\alph{subfigure})}
\renewcommand{\thefigure}{\Alph{section}\arabic{figure}}
\makeatletter \def\p@subfigure{\thefigure} \makeatother 

\section{Results in the laboratory frame}
\label{sec-appendix-lab-frame}

In this appendix we show laboratory-frame versions of some of our results for comparison with their rotating-frame counterparts; we include them separately to avoid the chance of confusion between the two frames. In the laboratory frame, the SFA formalism is well known \cite{LewensteinHHG, HHGTutorial}, and in essence it requires the calculation of
{
\abovedisplayskip=0.25em
\belowdisplayskip=0.25em
\begin{align}
\vbD_\lab (\Omega)
& = 
i
\int_{-\infty}^\infty\!\!\! \d t \!\!
\int_{\tref}^t \!\!\! \d t' \,
\vbd^*
  \mathopen{}\left(
  \vbp_{\lab,s}(t,t')+\vba_\lab(t)
  \right)\mathclose{}
\,
\nonumber \\ & \quad\times 
\left(\frac{2\pi}{\eps+i(t-t')}\right)^{3/2}
\Y\mathopen{}\left(
  \vbp_{\lab,s}(t,t')+\vba_\lab(t'))
\right)\mathclose{}
\nonumber \\ & \quad \times 
e^{
   +i\Omega t
   -iI_p(t-t')
   -\frac i2 \int^t_{t'} (\vbp_{\lab,s}(t,t') + \vba_\lab(\tau))^2 \d \tau
   }
.
\label{harmonic-dipole-lab-frame}
\end{align}
}

\onecolumngrid

\begin{figure*}[b!h]
\vspace{4mm}
\centering
\setlength{\tabcolsep}{0.3mm}
\begin{tabular}{cc}
\subfigure{\label{fig-lab-frame-harmonic-spectrum-p-}%
\includegraphics[scale=1]{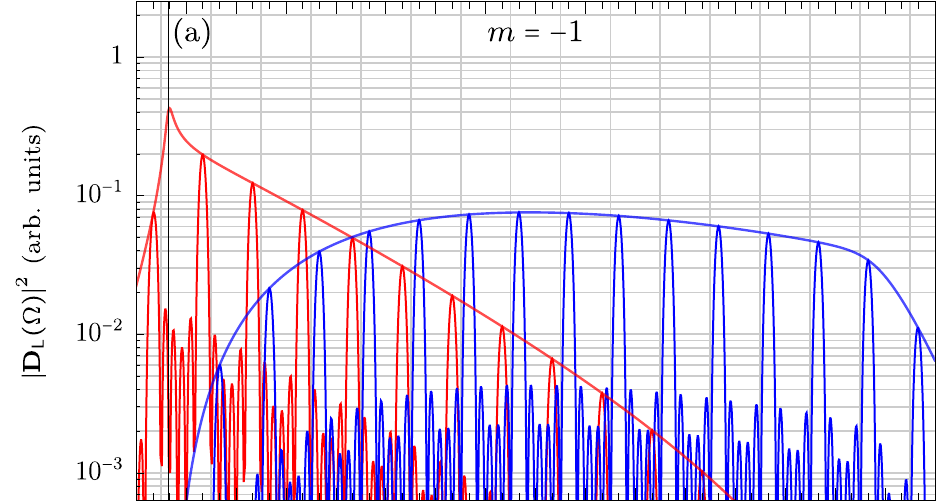}}&
\subfigure{\label{fig-lab-frame-harmonic-spectrum-p+}%
\includegraphics[scale=1]{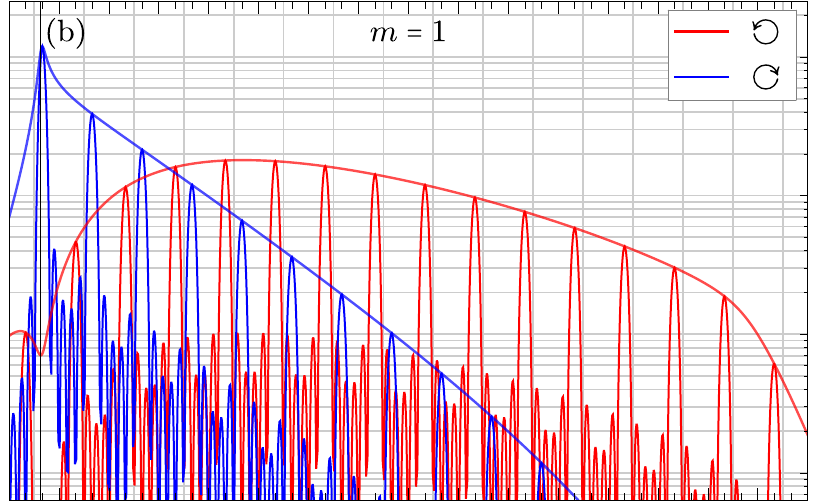}}\\[-1.4mm]
\includegraphics[scale=1]{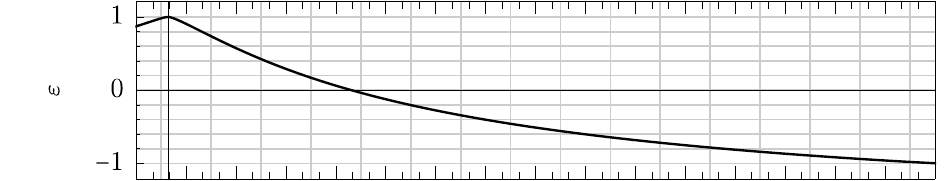}&
\includegraphics[scale=1]{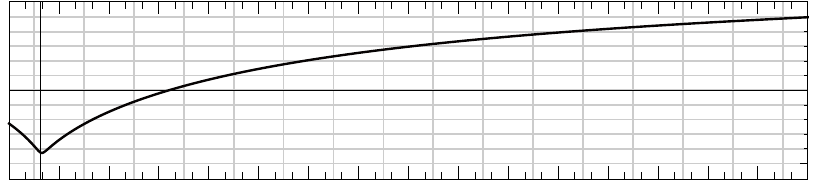}\\[-0.5mm]
\subfigure{\label{fig-lab-frame-harmonic-spectrum-s}%
\includegraphics[scale=1]{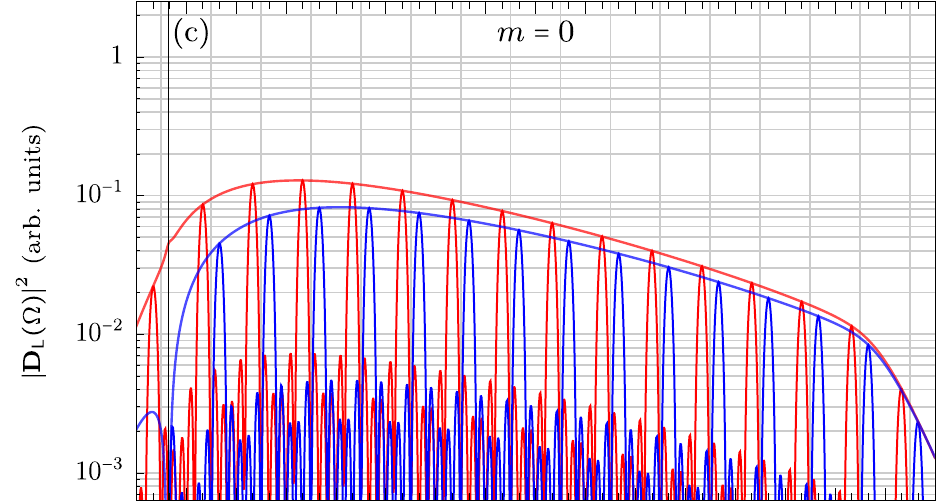}}&
\subfigure{\label{fig-lab-frame-harmonic-spectrum-2p}%
\includegraphics[scale=1]{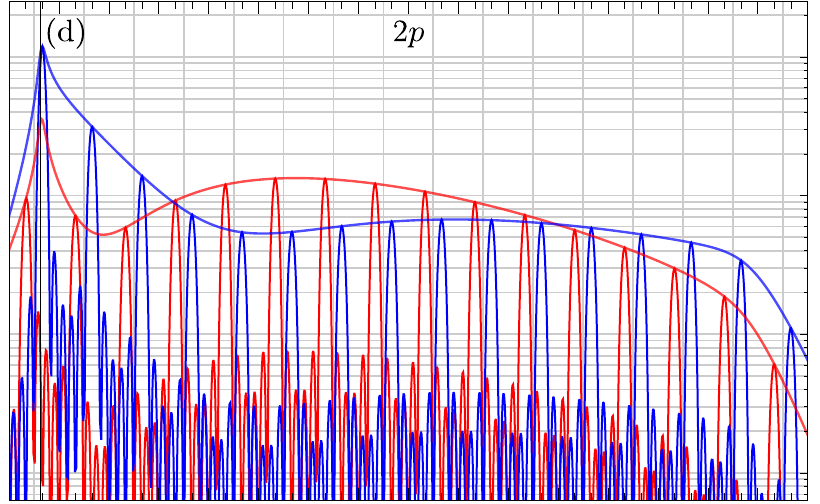}}\\[-1.4mm]
\includegraphics[scale=1]{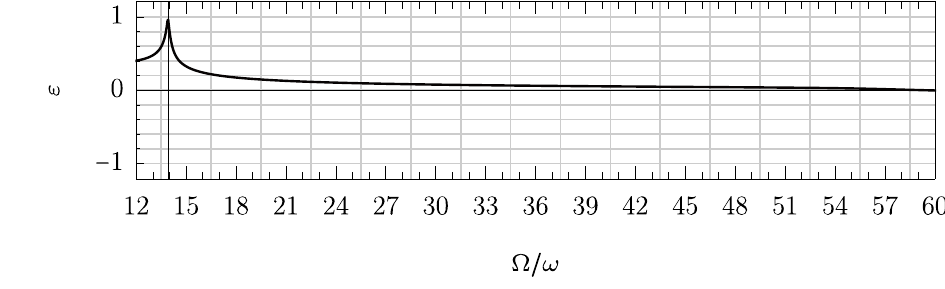}&
\includegraphics[scale=1]{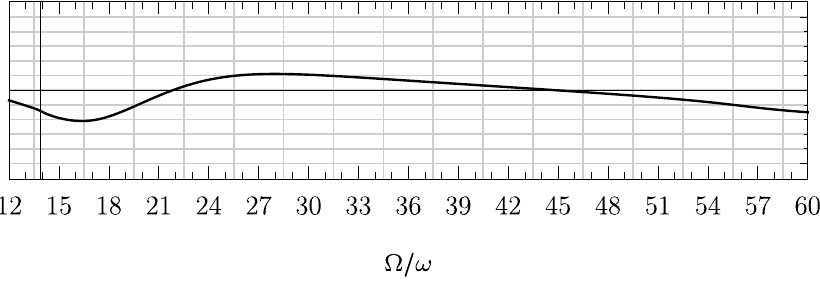}
\end{tabular}
\caption{
Harmonic spectra in the laboratory frame, presented as in \reffig{fig-harmonic-spectra-rotating-frame}. 
}
\label{fig-harmonic-spectra-lab-frame}
\end{figure*}

\begin{figure*}[t!]
\centering
\setlength{\tabcolsep}{-0.2mm}
\begin{tabular}{ccc}
\subfigure{ \label{fig-lab-frame-recombination-dipole-p-}
  \includegraphics[scale=1]{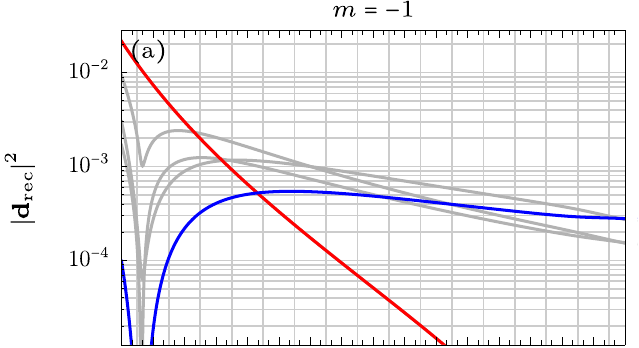}}&
\subfigure{ \label{fig-lab-frame-recombination-dipole-s}
  \includegraphics[scale=1]{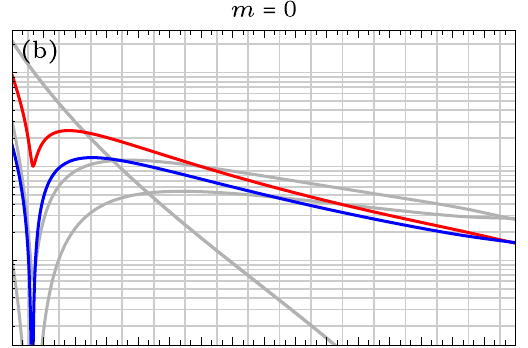}}&
\subfigure{ \label{fig-lab-frame-recombination-dipole-p+}
  \includegraphics[scale=1]{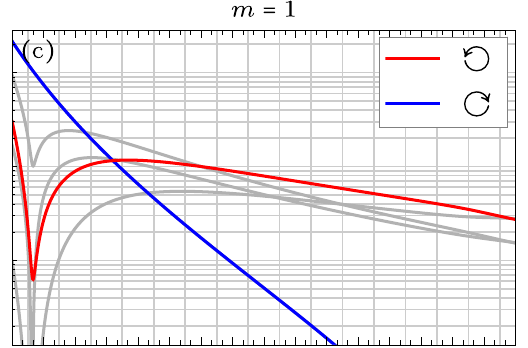}}\\[-3mm]
\subfigure{ \label{fig-lab-frame-ionization-factor-p-}
  \includegraphics[scale=1]{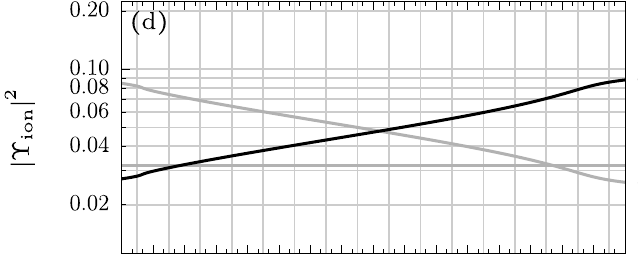}}&
\subfigure{ \label{fig-lab-frame-ionization-factor-s}
  \includegraphics[scale=1]{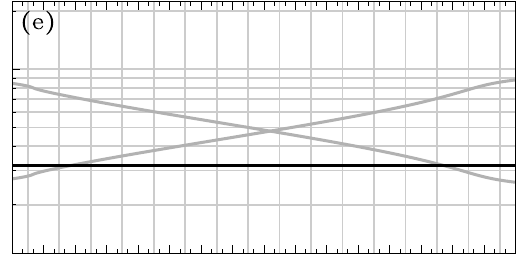}}&
\subfigure{ \label{fig-lab-frame-ionization-factor-p+}
  \includegraphics[scale=1]{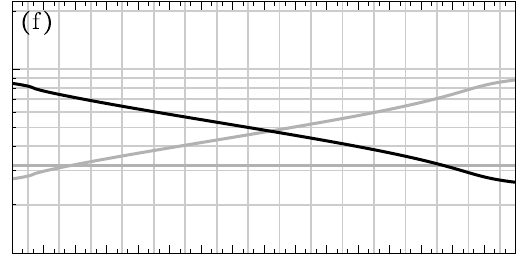}}\\[-3mm]
\subfigure{ \label{fig-lab-frame-action-factor-p-}
  \includegraphics[scale=1]{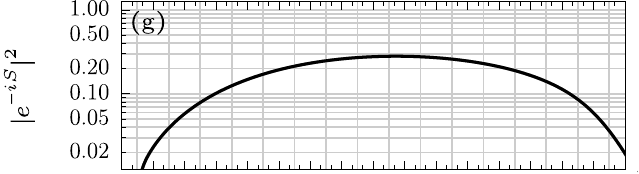}}&
\subfigure{ \label{fig-lab-frame-action-factor-s}
  \includegraphics[scale=1]{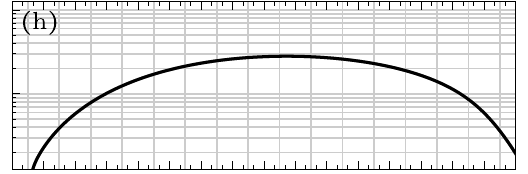}}&
\subfigure{ \label{fig-lab-frame-action-factor-p+}
  \includegraphics[scale=1]{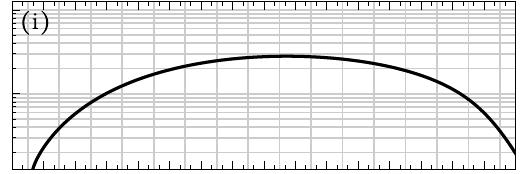}}\\[-3mm]
\subfigure{ \label{fig-lab-frame-total-factor-p-}
  \includegraphics[scale=1]{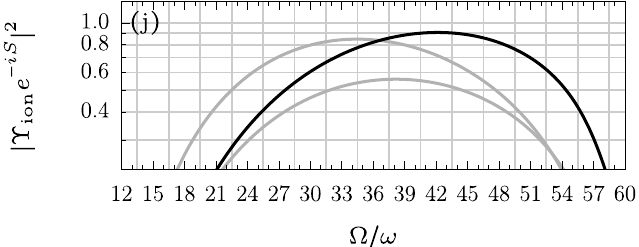}}&
\subfigure{ \label{fig-lab-frame-total-factor-s}
  \includegraphics[scale=1]{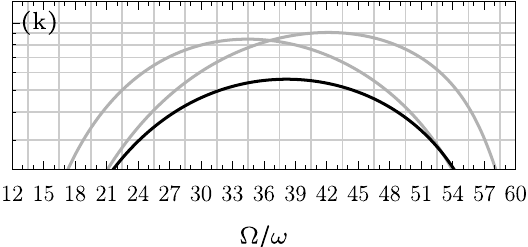}}&
\subfigure{ \label{fig-lab-frame-total-factor-p+}
  \includegraphics[scale=1]{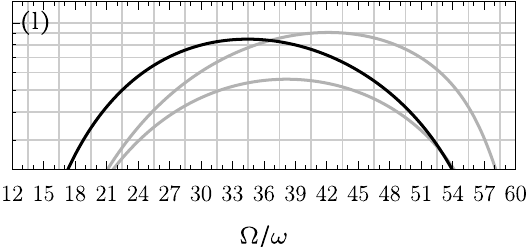}}
\end{tabular}
\caption{
Geometrical factors affecting the helicity asymmetry and the harmonic intensity for the different orbitals in the laboratory frame, presented as in \reffig{fig-internal-structure-of-spectra}.
}
\label{fig-internal-structure-of-spectra-lab-frame}
\end{figure*}

%
\twocolumngrid

We show the resulting harmonic spectra in \reffig{fig-harmonic-spectra-lab-frame}, which exhibits the usual selection rules, with right-handed harmonics at $(3n+1)\omega$ and left-handed harmonics at $(3n+2)\omega$. The $s$-state emission, shown in \reffig{fig-lab-frame-harmonic-spectrum-s}, shows the usual helicity asymmetry, while the total $p$-state emission in \reffig{fig-lab-frame-harmonic-spectrum-2p} has an enhanced asymmetry in the mid-plateau, which is then lost near the cutoff.

Similarly, we show in \reffig{fig-internal-structure-of-spectra-lab-frame} the internal structure of the factors that lead to the helicity asymmetry, as in \reffig{fig-internal-structure-of-spectra}; here the action is independent of $m$, and the reduced ionization matrix elements $|\Y_\mathrm{ion}|^2$ in Figs.~\ref{fig-internal-structure-of-spectra-lab-frame}(\hyperref[fig-lab-frame-ionization-factor-p-]{d}-\hyperref[fig-lab-frame-ionization-factor-p+]{f}) do not show the shifts they exhibit in the rotating frame, so that the total ionization factor $|\Y_\mathrm{ion} e^{-iS}|^2$ in Figs.~\ref{fig-internal-structure-of-spectra-lab-frame}(\hyperref[fig-lab-frame-total-factor-p-]{j}-\hyperref[fig-lab-frame-total-factor-p+]{l}) is essentially identical to that in the rotating frame.

\begin{figure*}[t!]
\centering
\setlength{\tabcolsep}{0.1mm}
\begin{tabular}{rl}
\subfigure{ \label{fig-lab-frame-trajectory-position-17}
  \includegraphics[scale=1]{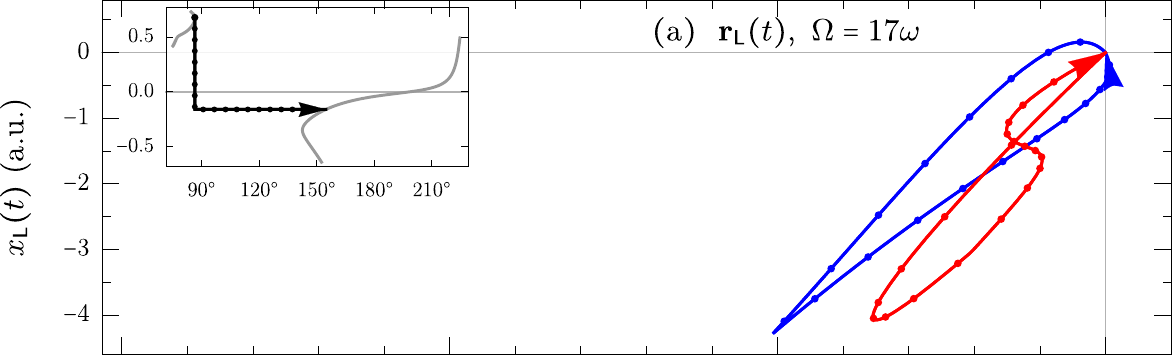}}&
\subfigure{ \label{fig-lab-frame-trajectory-velocity-17}
  \includegraphics[scale=1]{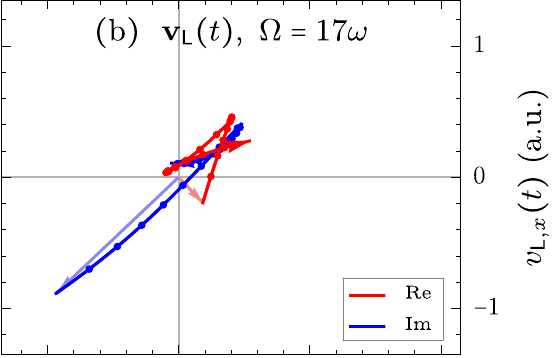}}\\[-2.7mm]
\subfigure{ \label{fig-lab-frame-trajectory-position-27}
  \includegraphics[scale=1]{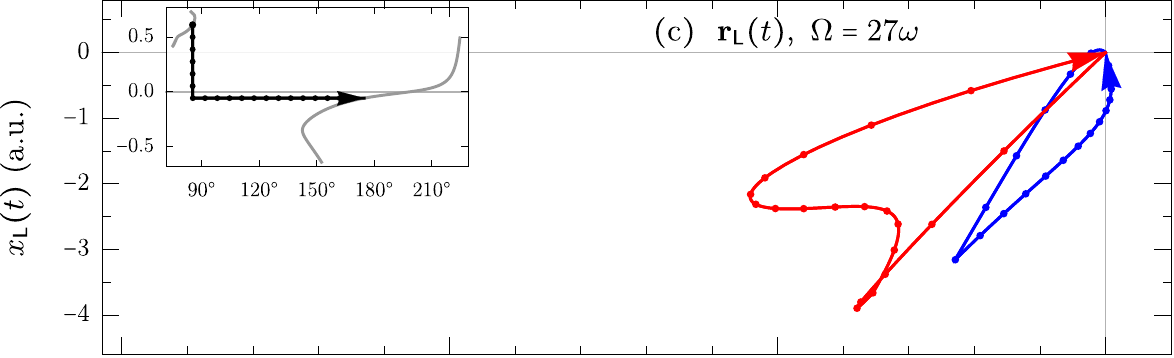}}&
\subfigure{ \label{fig-lab-frame-trajectory-velocity-27}
  \includegraphics[scale=1]{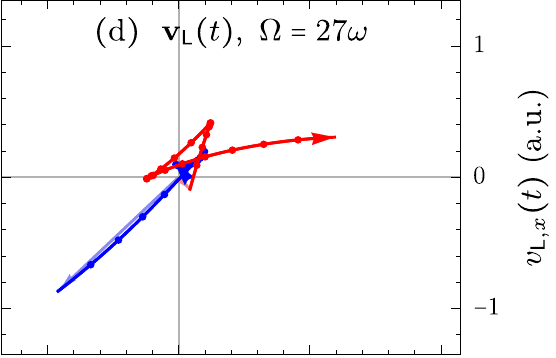}}\\[-2.7mm]
\subfigure{ \label{fig-lab-frame-trajectory-position-37}
  \includegraphics[scale=1]{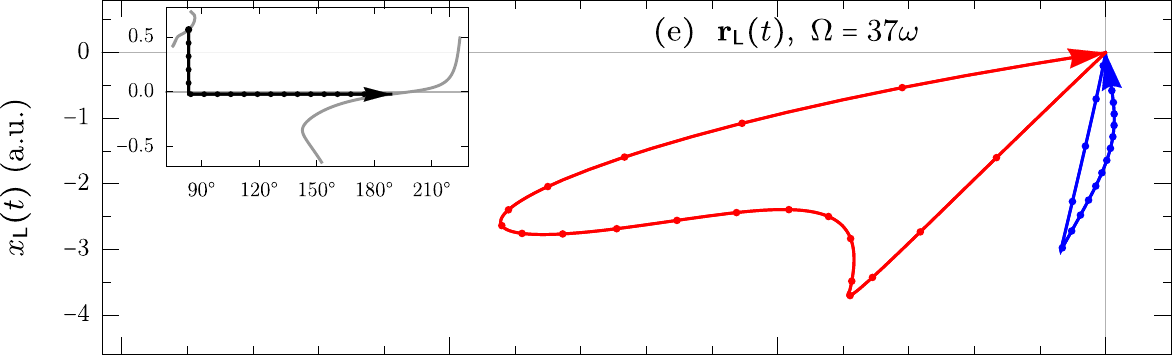}}&
\subfigure{ \label{fig-lab-frame-trajectory-velocity-37}
  \includegraphics[scale=1]{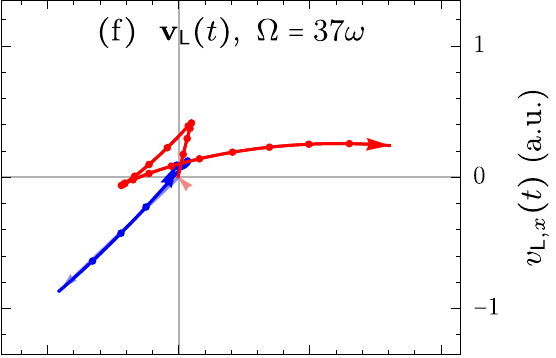}}\\[-2.7mm]
\subfigure{ \label{fig-lab-frame-trajectory-position-47}
  \includegraphics[scale=1]{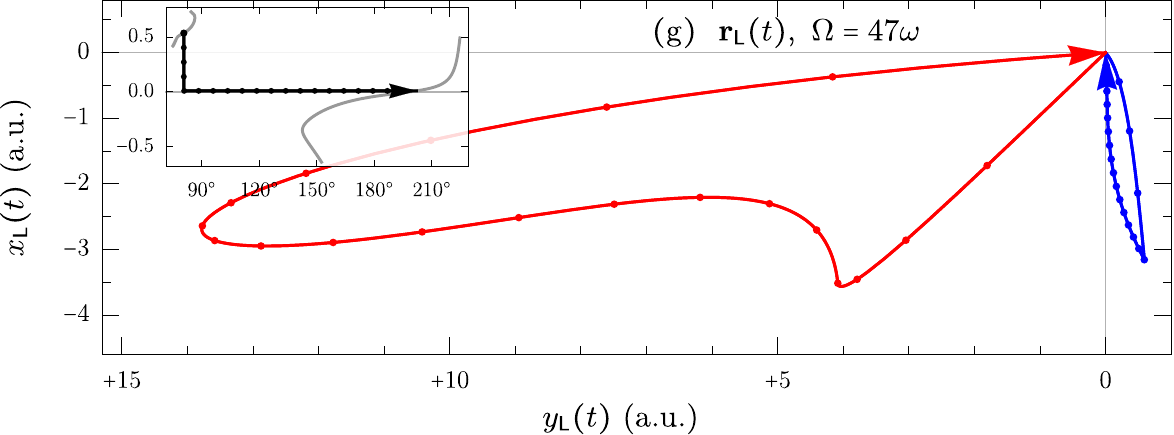}}&
\subfigure{ \label{fig-lab-frame-trajectory-velocity-47}
  \includegraphics[scale=1]{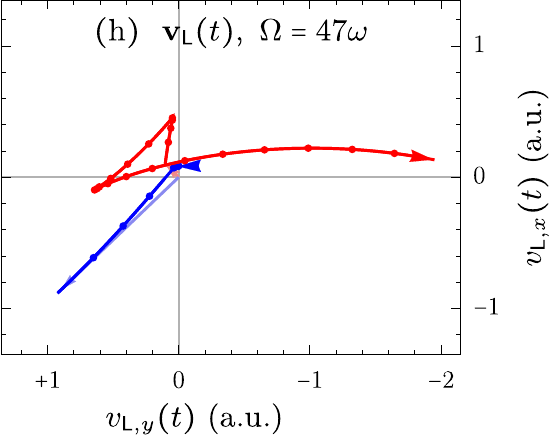}}
\end{tabular}
\caption{
Photoelectron trajectories in the laboratory frame, presented as in \reffig{fig-trajectories}. To aid comparison, we have applied a global rotation by $\SI{90}{\degree}$ counterclockwise, as shown in the tick marks and axis labels, which does not otherwise affect the results.
}
\label{fig-trajectories-lab-frame}
\end{figure*}

Finally, we present in \reffig{fig-trajectories-lab-frame} the laboratory-frame quantum-orbit trajectories that correspond to the same situations as in \reffig{fig-trajectories}, using the same conventions. In contrast to the rotating frame, the ionization velocity has an imaginary part that can be in either chiral relation to the real part: while $\Im(\vb v_\lab(t_\mathrm{i}))$ remains in the fourth quadrant (pale blue arrow), $\Re(\vb v_\lab(t_\mathrm{i}))$ changes from the second to the fourth quadrants (pale red arrow), going through a zero at $\Omega\approx 37\omega$, which corresponds to the crossing between the $p$-state ionization matrix elements in Figs.~\ref{fig-internal-structure-of-spectra-lab-frame}(\hyperref[fig-lab-frame-ionization-factor-p-]{d}-\hyperref[fig-lab-frame-ionization-factor-p+]{f}). 

In terms of the position-space trajectories, this helicity crossing at $\Omega\approx 37\omega$ can be seen in the shift of the imaginary part of the trajectory as the photoelectron departs from the origin during the tunnelling step: before the crossing $\Im(\vbr_\lab(t))$ is clockwise from $\Re(\vbr_\lab(t))$, and after the crossing it lies counter-clockwise from it. At the crossing itself, the real part of the ionization velocity is exactly zero, which means that $\Im(\vbr_\lab(t))$ is quadratic, instead of linear, in $t-t_\mathrm{i}$ at the moment of ionization, with the imaginary trajectory initially lying along the real part of the position.

\section{Transition dipoles for \textit{p} states}
\label{sec-appendix-matrix-elements}
In this appendix we calculate the functions $\vbd(\vbk)=\matrixel{\vbk}{\hat{\vbr}}{g}$ and $\Psi(\vbk) = \braket{\vbk}{g}$ for the ground states of interest, the $2p_\pm$ and a (fictional) $s$ states of neon. We model these using a short-range potential, which gives a ground-state wavefunction of the form
\begin{equation}
\braket{\vbr}{\varphi_{lm}}
=Y_{lm}(\hat\vbr)\varphi_{lm}(r)
=Y_{lm}(\hat\vbr)  C\frac{e^{-\kappa r}}{\kappa r}
,
\label{short-range-position-wavefunction}
\end{equation}
where $\frac12\kappa^2 = I_p$. We begin with the simpler quantity, the momentum-space wavefunction, calculating the inner product over position space. To do this, we separate the plane-wave factor into partial waves~\cite{JacksonElectrodynamics}, in the form
\begin{align}
e^{i\vbk\cdot\vbr}
& =
4\pi
\sum_{l'=0}^\infty
\sum_{m'=-l'}^{l'}
i^{l'}
j_{l'}(kr)
Y_{l'm'}(\hat\vbr)
Y_{l'm'}^*(\hat\vbk)
\nonumber \\ & =
4\pi
\sum_{l'=0}^\infty
\sum_{m'=-l'}^{l'}
i^{l'}
\frac{j_{l'}(kr)}{k^{l'}}
Y_{l'm'}(\hat\vbr)
S_{l'm'}^*(\vbk)
,
\end{align}
where we turn the momentum spherical harmonic into a solid harmonic $S_{lm}(\vbk)=k^{l}Y_{lm}(\hat\vbk)$, a homogeneous polynomial of degree $l$ in the Cartesian components of $\vbk$, since we are interested in maintaining explicit analyticity with respect to those harmonics. (Similarly, the Bessel factor $j_{l'}(kr)/k^{l'}$ is guaranteed to be an entire function of $k^2$ because of the low-argument asymptotics $j_{l'}(kr)\sim(kr)^{l'}$.)

\subsection{Momentum-space wavefunctions}
Using this decomposition, we can express $\Psi(\vbk)$ in the form
{
\allowdisplaybreaks
\begin{align}
\Psi_{lm}(\vbk) 
& = 
\braket{\vbk}{\varphi_{lm}}
=
\int \frac{\d\vbr}{(2\pi)^{3/2}\!\!}\,\,
e^{-i\vbk\cdot\vbr}
\varphi_{lm}(\vbr)
\nonumber \\ & =
\sum_{l'=0}^\infty
\sum_{m'=-l'}^{l'}
\frac{2i^{-l'}}{\sqrt{2\pi}}
S_{l'm'}(\vbk)
\int
Y_{l'm'}^*(\hat\vbr)
Y_{lm}(\hat\vbr)
\d\Omega
\nonumber \\ & \qquad \times
\int_0^\infty 
\frac{1}{k^{l'}}
j_{l'}(kr)
\varphi_{lm}(r)
r^2
\d r
\nonumber \\ & =
\frac{2i^{-l}}{\sqrt{2\pi}}
S_{lm}(\vbk)
\int_0^\infty 
\frac{1}{k^{l}}
j_{l}(kr)
\varphi_{lm}(r)
r^2
\d r
\nonumber \\ & = 
S_{lm}(\vbk)
G_{l}(k)
,
\end{align}
}
where the angular integral reduces to $\delta_{ll'}\delta_{mm'}$, giving only a single term in $\Psi$ with the same angular dependence as $\braket{\vbk}{\varphi_{lm}}$, and a radial term which we encapsulate into
\begin{align}
G_{l}(k)
& = 
\frac{2i^{-l}}{\sqrt{2\pi}}
\int_0^\infty 
\frac{1}{k^{l}}
j_{l}(kr)
\varphi_{lm}(r)
r^2
\d r
,
\end{align}
an analytic function of $k^2$. To calculate this, we now need to put in an explicit ground-state wavefunction.

The simplest is the $2s$ state, which has the angular dependance $S_{00}(\vbk)=1/\sqrt{4\pi}$, and for which the radial integral reduces to
\begin{align}
G_{0}(k)
& = 
\frac{2}{\sqrt{2\pi}}
\frac{C}{\kappa}
\int_0^\infty 
j_{0}(kr)
e^{-\kappa r}
r
\d r
%
= 
\frac{2}{\sqrt{2\pi}}
\frac{C}{\kappa}
\frac{1}{k^2+\kappa^2}
,
\end{align}
giving
\begin{align}
\Psi_{00}(\vbk) 
& = 
\frac{C/\kappa}{\sqrt{2}\pi}
\frac{1}{k^2+\kappa^2}
.
\end{align}
This is an analytical function of $k^2$, and it has a pole at $k=i\kappa$ which is then regularized by the passage to $\Y_{00}(\vbk)=\frac12(k^2+\kappa^2)\Psi_{00}(\vbk)$, giving a fully regular integrand at the ionization saddle point.

\begin{figure}[htb]
\centering
\includegraphics[scale=1]{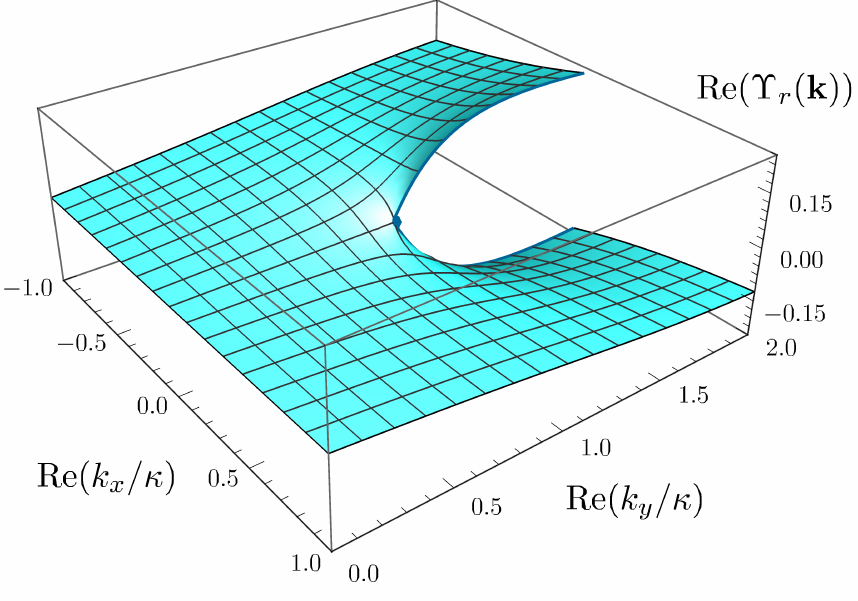}
\caption{Regularized radial dependence of the momentum-space wavefunction for a $2p$ state, $\Y_r(\vbk) = \frac12(k^2+\kappa^2)G_1(k)$, over the complex $k_x$ plane with $k_y=k_z=0$, showing a branch cut at the ionization momentum, $k_x=i\kappa$.}
\label{fig:upsilon-branch-cuts}
\end{figure}

For $2p$ states, the situation is slightly more complicated, although the radial integral
\begin{align}
G_{1}(k)
& = 
\frac{2i^{-1}}{\sqrt{2\pi}}
\frac{C}{\kappa k^l}
\int_0^\infty 
j_{1}(kr)
e^{-\kappa r}
r
\d r
\nonumber \\ & =
\frac{2iC}{\sqrt{2\pi}\kappa}
\frac{1}{k^2}
\left[
\frac{\kappa}{k^2+\kappa^2}
-
\frac{1}{k}
\arctan(\frac{k}{\kappa})
\right]
.
\label{p-state-radial-G}
\end{align}
is similarly easy to integrate. The result, however, offers some nontrivial subtleties, since it appears to have a pole at $k=0$. This is fortunately a mirage, since the constant parts of the two terms in the numerator cancel out, and the radial integral has a small-$k$ expansion of the form
\begin{align}
G_{1}(k)
& = 
\frac{2iC}{\sqrt{2\pi}}
\left[
-\frac{2}{3\kappa^4}
+\frac{4k^2}{5\kappa^6}
-\frac{6k^4}{7\kappa^8}
+O(k^6/\kappa^{10})
\right]
,
\end{align}
with no singular terms. On the other hand, the radial integral in~\eqref{p-state-radial-G} does suffer from a more serious problem, in the form of a branch cut at $k=i\kappa$. This branch cut comes from the behaviour of the $\arctan$ term at complex arguments, and it ultimately derives from the form of the general result for $G_{l}(k)$,
\begin{align}
G_{l}(k)
& = 
\frac{\Gamma(l+2)}{\Gamma(l+3/2)}
\frac{
  C
  }{
  2^{l+\frac12}
  i^{l}
  \kappa^{l+3}
}
\nonumber \\ & \qquad \times
{}_2F_1\mathopen{}\left(
  \frac{l}{2}+1,  \frac{l}{2}+\frac{3}{2};  l+\frac{3}{2};  -\frac{k^2}{\kappa^2}
\right)\mathclose{}
,
\end{align}
which has a natural branch cut at $\frac{k^2}{\kappa^2}=-1$ that only vanishes at $l=0$. We show this branch cut for the $l=1$ case in \reffig{fig:upsilon-branch-cuts}. In addition to the branch cut, the momentum-space wavefunction $\Psi_{2m}(\vbk)$ is actually singular at the branch point, but as with $\Psi_{00}(\vbk)$ this singularity gets regularized in the passage to $\Y_{2m}(\vbk)=\frac12(k^2+\kappa^2)\Psi_{2m}(\vbk)$.

Unfortunately, the branch cut is in an inconvenient location, because the requirement that $k^2+\kappa^2=0$ is precisely the saddle-point equation for the ionization time for the $2s$ state. This point is not problematic, but the saddle point for the $2p_-$ state gets pushed to higher imaginary part due to its increased ionization potential in the rotating frame, and it would therefore fall on or near the branch cut. This means, therefore, that for consistency we will need to retain the saddle point for the $2s$ state for use in the ionization matrix element $\Y\mathopen{}\left(\vbp_{\rot,s}(t,t')+\vba_\rot(t')) \right)\mathclose{}$.

It is also worth remarking that this branch cut, along with the singularities in $\Psi_{00}(\vbk)$ and $\Psi_{2m}(\vbk)$ at $k=i\kappa$, are natural features of the momentum-space wavefunction, and they can already be discerned from its definition,
\begin{align}
\Psi_{lm}(\vbk) 
& = 
\frac{C}{(2\pi)^{3/2}\!\!}
\int \,\,
\frac{Y_{lm}(\hat\vbr)}{\kappa r}
e^{-i\vbk\cdot\vbr - \kappa r}
\d\vbr
.
\end{align}
Here, for real $\vbk$, the integral is strongly confined by the radial exponential $e^{-\kappa r}$, and adding some imaginary parts to the Fourier kernel does not push the envelope too far. However, a momentum of the form $\vbk =i\kappa \ue{x}$ will change the Fourier kernel into an exponential $e^{+\kappa x}$ which is no longer well controlled by the radial exponential, so there is no longer a guarantee of a convergent integral.

\subsection{Dipole transition matrix elements}
The considerations for the dipole transition matrix elements $\vbd(\vbk)=\matrixel{\vbk}{\hat{\vbr}}{g}$ are similar to the above, but the presence of the vector operator introduces some additional complexity. Thus, we can use the same partial-wave expansion for the plane wave as above, but we also need to use appropriate language for the vector dipole operator, which we decompose in the form
\begin{align}
\label{r-in-circular-basis}
\vbr 
& =
 r_x \hat{\mathbf{e}}_x + r_y \hat{\mathbf{e}}_y + r_z \hat{\mathbf{e}}_z 
\nonumber \\ & =
\frac{1}{\sqrt{2}} (-r_x + i r_y) \hat{\mathbf{e}}_1 + \frac{1}{\sqrt{2}} (r_x + i r_y) \hat{\mathbf{e}}_{-1} + r_z \hat{\mathbf{e}}_0,
\end{align}
or, equivalently,
\begin{equation}
\vbr
=
\sqrt{\frac{4\pi}{3}}
r
\sum_{q=-1}^{1}
Y_{1q}(\hat{\vbr})
\ue{q}^*
,
\label{r-as-sum-of-Ylm}
\end{equation}
where we have used the definition of the circular basis $\hat{\mathbf{e}}_{\pm} = \mp (\hat{\mathbf{e}}_x \pm i \hat{\mathbf{e}}_y)/\sqrt{2}$.
This then lets us decompose the matrix element, which originally reads
\begin{align}
\vbd_{lm}^*(\vbk) 
& = 
\matrixel{\varphi_{lm}}{\hat\vbr}{\vbk}
%
=
\int \frac{\d\vbr}{(2\pi)^{3/2}\!\!}\,\,
e^{+i\vbk\cdot\vbr}
\, \vbr \, 
\varphi_{lm}^*(\vbr)
,
\label{full-expansion-d-k-initial}
\end{align}
in the form
\begin{align}
\vbd_{lm}^*(\vbk) 
& = 
\sum_{q=-1}^{1}
\sum_{l'=0}^\infty
\sum_{m'=-l'}^{l'}
\frac{2^{3/2}i^{l'}}{\sqrt{3}}
\ue{q}^*
S_{l'm'}(\vbk)
\nonumber \\ & \qquad \qquad \qquad \qquad \times
\int
Y_{l'm'}^*(\hat\vbr)
Y_{1q}(\hat{\vbr})
Y_{lm}^*(\hat\vbr)
\d\Omega
\nonumber \\ & \qquad \qquad \qquad \qquad \times
\int_0^\infty 
\frac{1}{k^{l'}}
j_{l'}(kr)
\varphi_{lm}(r)
r^3
\d r
.
\label{full-expansion-d-k}
\end{align}
%
%
We focus from this early stage on the complex conjugate $\vbd_{lm}^*(\vbk)  = \matrixel{\varphi_{lm}}{\hat\vbr}{\vbk}$ of the transition dipole, which needs to be calculated explicitly for the global temporal integrand to be analytical.

For the dipole, the key change with respect to the momentum-space wavefunction is in the angular integral, which changes from an inner product to a triple product of spherical harmonics. This can still be handled easily~\citenisteq{34.3.22}, and it evaluates to a product of Wigner $3j$ symbols,
\begin{align}
\int
&
Y_{l'm'}^*(\hat\vbr)
Y_{1q}(\hat{\vbr})
Y_{lm}^*(\hat\vbr)
\d\Omega
\nonumber \\ & =
(-1)^{m+m'}\!\!\!\!
\int
Y_{l',-m'}(\hat\vbr)
Y_{1q}(\hat{\vbr})
Y_{l,-m}(\hat\vbr)
\d\Omega
\\ & = 
(-1)^{m+m'}
\sqrt{\frac{3(2l+1)(2l'+1)}{4\pi}}
\nonumber \\ & \qquad \times
\begin{pmatrix}
l & l' & 1 \\ 0 & 0 & 0
\end{pmatrix}
\begin{pmatrix}
l & l' & 1 \\ -m & -m' & q
\end{pmatrix}
.
\end{align}
As above, this gives a restriction on the partial waves which can contribute to the final expression: we require that $m'=q-m$, $l+l'+1$ needs to be even, and every combination of $l$, $l'$ and $1$ needs to obey the triangle inequalities. In particular, this means that we require $|l-l'|\leq 1$ and $l+l'$ to be odd, so only $l'=l \pm 1$ can contribute.

This then lets us cut down substantially on the form of our matrix element, which can be expressed as
\begin{align}
\vbd_{lm}^*(\vbk) 
& = 
\sum_{q=-1}^{1}
\sum_{l'=l\pm1}
\ue{q}^*
N_{ll'mq}
S_{l',q-m}(\vbk)
F_{l'l}(k)
, \  \text{where}
\\
N_{ll'mq}
& =
2^{3/2}i^{l'}
(-1)^{q}
\sqrt{\frac{(2l+1)(2l'+1)}{4\pi}}
\begin{pmatrix}
l & l' & 1 \\ 0 & 0 & 0
\end{pmatrix}
\nonumber \\ & \qquad \times
\begin{pmatrix}
l & l' & 1 \\ -m & m-q & q
\end{pmatrix}
\label{constant-Nllmq-definition}
\ \text{and}
\\
F_{l'l}(k)
& =
\int_0^\infty 
\!
\frac{1}{k^{l'}}
j_{l'}(kr)
\varphi_{lm}(r)
r^3
\d r
.
\label{first-full-breakout}
\end{align}
Here $F_{l'l}(k)$ now carries the bulk of the computation, but for the short-range wavefunction of~\eqref{short-range-position-wavefunction} it can be integrated exactly even in the general case~\cite[Eq.~(6.621.1)]{gradshteyn_ryzhik}:
\begingroup
\allowdisplaybreaks
\begin{align}
F_{l'l}(k)
& =
\frac{C/\kappa}{k^{l'}}
\int_0^\infty 
j_{l'}(kr)
e^{-\kappa r}
r^2
\d r
\nonumber \\ & =
\frac{C/\kappa}{ k^{l'+1/2}}
\sqrt{\frac{\pi}{2}}
\int_0^\infty 
J_{l'+1/2}(kr)
e^{-\kappa r}
r^{3/2}
\d r
\nonumber \\ & =
\frac{\sqrt{\pi}C}{2^{l'+1} \kappa^{l'+4}}
\frac{\Gamma(l'+3)}{\Gamma(l'+3/2)}
\nonumber \\ & \qquad \times
{}_2F_1\mathopen{}\left(
  \frac{l'+3}{2},
  \frac{l'}{2}+2;
  l'+\frac{3}{2};
  -\frac{k^2}{\kappa^2}
  \right)\mathclose{}
.
\end{align}
\endgroup

The forms above for the transition dipole are general, but for the case of the $s$ states they overestimate the complexity of the required expressions. In this case, we have only one intermediate momentum contributing, $l'=1$, and therefore only one integral to consider, which reduces to
\begin{align}
F_{10}(k)
& =
\frac{2C/\kappa}{(k^2+\kappa^2)^2}
,
\end{align}
with the constant similarly reducing to $N_{010q}=i\sqrt{2/3\pi}$, so the sum over $q$ reduces to the same vector sum as in~\eqref{r-as-sum-of-Ylm}, so we have
\begin{align}
\vbd^*\!(\vbk) 
& = 
i\sqrt{\frac{2}{3\pi}}
F_{10}(k)
\sum_{q=-1}^{1}
\,\,
\ue{q}^*
S_{1,q}(\vbk)
\nonumber \\ & =
\frac{i\sqrt{2}C}{\pi\kappa}
\frac{
  \vbk
  }{
  (k^2+\kappa^2)^2
  }
.
\label{s-state-transition-dipole-final}
\end{align}
As expected for a spherically symmetric state, the transition dipole points exclusively in the direction of the plane wave's momentum.

For the $2p$ states, the situation is more complicated, partly because we now have two integrals to calculate, which come down to
\begin{align}
F_{01}(k)
& =
\frac{2C}{(k^2+\kappa^2)^2}
\label{radial-f20-result}
\quad \text{and} 
\\
F_{21}(k)
& =
\frac{C}{k^4}
\left[
 -
 \frac{3\kappa^2+5k^2}{(k^2+\kappa^2)^2}
 +
 \frac{3}{k\kappa}
 \arctan(\frac{k}{\kappa}) 
\right]
.
\label{radial-f21-result}
\end{align}
Here $F_{21}(k)$ shares many of the same features as $F_{10}(k)$ as discussed above. For one, it appears singular because of the factor $1/k^4$, but both the constant and $k^2$ terms of the two terms inside the square brackets cancel out, to give a global Taylor series which is regular and nonzero at the origin. Similarly, $F_{21}$ has a branch cut at $k=i\kappa$, and it has a singularity at that branch point, but in contrast with $\Psi(\vbk)$ this pole is now of second order, so it would not be regularized by adding a factor of $\frac12(k^2+\kappa^2)$. Fortunately, this is not a problem since we will only need this transition dipole at the recollision momentum, for which $\frac12(k^2+\kappa^2) = \Omega$, the harmonic photon energy, is real and positive.

To obtain the transition dipole, we now put all of this together. We knew already that 
\begin{align}
\vbd_{lm}^*(\vbk) 
& = 
\sum_{q=-1}^{1}
\sum_{l'=l\pm1}
N_{ll'mq}
\ue{q}^*
S_{l',q-m}(\vbk)
F_{l'l}(k)
\end{align}
and we can further simplify things since the $l'=0$ electronic monopole term requires a polarization along $q=m$, giving us
\begin{align}
\vbd_{1m}^*(\vbk) 
& {=} 
\ue{m}^*
\left(
N_{10mm}
S_{00}(\vbk)
F_{01}(k)
{+}
N_{12mm}
S_{20}(\vbk)
F_{21}(k)
\right)
\nonumber \\ & \quad + 
\ue{-m}^*
N_{12m,-m}
S_{2,-2m}(\vbk)
F_{21}(k)
\nonumber \\ & \quad
+ 
\ue{0}^*
N_{12m0}
S_{2,-m}(\vbk)
F_{21}(k)
.
\label{p-state-transition-dipole-final}
\end{align}
Here the $F_{l'l}(k)$ are as in~\eqref{radial-f20-result} and~\eqref{radial-f21-result}, the $N_{ll'mq}$ are given by~\eqref{constant-Nllmq-definition}, and the solid harmonics are given by
\begin{subequations}
\begin{align}
S_{00}(\vbk)
& = 
\frac{1}{\sqrt{4\pi}},
\\
S_{20}(\vbk)
& = 
\sqrt{\frac{5}{16\pi}}
\left( 2k_z^2 -k_x^2 -k_y^2 \right),
\ \text{and } 
\\
S_{2,-2m}(\vbk)
& = 
\sqrt{\frac{5}{32\pi}}
\left( k_x - imk_y \right)^2
.
\end{align}
\end{subequations}
There is also in \eqref{p-state-transition-dipole-final} a contribution along $\ue0=\ue z$, for which the solid harmonics are given by $S_{2,-m}(\vbk) = m \sqrt{\frac{15}{16\pi}}\left( k_x - imk_y \right)k_z$, and which does not contribute to the harmonic emission in problems confined to the $x,y$ plane; we nevertheless include it for completeness.

In this form, the recombination dipole looks fairly complex, but its form in \eqref{p-state-transition-dipole-final} belies some of its underlying simplicity. More specifically, it is important to note that several of its components -- $F_{01}(k)$, $F_{21}(k)$ and $S_{20}(\vbk)$ -- are only functions of $k^2$ on the $x,y$ plane, and that they are therefore constrained by the recollision saddle-point equation,
\begin{equation}
\frac12 \vbk^2 + I_p = \Omega,
\label{simplified-recollision-saddle-point-equation}
\end{equation}
which forces them to be simple functions of the harmonic order, and that they are therefore insensitive to the details of the quantum-orbit dynamics.

In this connection, then, it is useful to write those functions, based on \eqref{simplified-recollision-saddle-point-equation}, as
\begingroup
\allowdisplaybreaks
\begin{align}
S_{20}(\vbk)
& =
\sqrt{\frac{5}{4\pi}}
(I_p-\Omega)
,
\\
F_{01}(k)
& =
\frac{C}{2} \frac{1}{\Omega^2}
, \quad \text{and} 
\\
F_{21}(k)
& =
\frac{C/4}{(\Omega-I_p)^2}
\left[
 \frac{2I_p-5\Omega}{2\Omega^2}
 \right. \\ & \qquad \left. \nonumber
 +
 \frac{3}{2I_p}
 \sqrt{\frac{I_p}{\Omega-I_p}}
 \arctan(\sqrt{\frac{\Omega-I_p}{I_p}}) 
\right]
,
\end{align}
\endgroup
where for high $\Omega$ the latter asymptotically approaches $F_{21}(k) \approx \frac{3\pi C}{8\sqrt{I_p}}(\Omega-I_p)^{-5/2}$, though that requires harmonic photon energies higher than $\Omega/I_p \gtrsim 10$, and the combination $S_{20}(\vbk)F_{21}(k)$ is mostly flat after its zero at threshold. This means, then, that we can write down an explicit expression for the harmonic component $\ue{-m}^*\cdot\vbd_{1m}^*(\vbk)$ which counter-rotates with respect to the ground state in terms of the harmonic photon energy
\begin{align}
\ue{-m}^*\cdot
\vbd_{1m}^*(\vbk) 
& = 
\frac{C}{4\sqrt{6}\,\pi}
\left[
\frac{2}{\Omega^2}
-
\frac{1}{\Omega-I_p}
\left(
 \frac{2I_p-5\Omega}{2\Omega^2}
 \right. \right. \nonumber \\ & \quad \left. \left.
 +
 \frac{3}{2I_p}
 \sqrt{\frac{I_p}{\Omega-I_p}}
 \arctan(\sqrt{\frac{\Omega-I_p}{I_p}}) 
\right)
\right]
,
\label{d-rec-factors-Omega-Ip}
\end{align}
which completely fixes its structure, and forces its steep decay throughout the plateau; the minor differences in this factor between Figs.~\ref{fig-recombination-dipole-p-} and \subref{fig-recombination-dipole-p-} are due to the effective Zeeman shifts, which change $I_p$ in \eqref{d-rec-factors-Omega-Ip} to $I_p-m\alpha$.

On the other hand, the component $\ue{m}^*\cdot\vbd_{1m}^*(\vbk)$ which co-rotates with the ground state is more complicated, because it depends on the solid harmonic $S_{2,-2m}(\vbk) = \sqrt{\frac{5}{8\pi}} k_{\pm}^2$ and therefore, as described in the text, is strongly affected by the quantum-orbit dynamics and by the passage to the rotating frame.

\bibliographystyle{arthur} 
\bibliography{references}{}


\end{document}